\documentclass[12pt]{spieman}  
\usepackage{amsmath,amsfonts,amssymb}
\usepackage{graphicx}
\usepackage{setspace}
\usepackage{tocloft}

\usepackage[separate-uncertainty=true]{siunitx}
\usepackage{tikz}
\usepackage{stmaryrd}
\usepackage{dirtytalk}
\usepackage[nomain,acronym]{glossaries}
\glsdisablehyper
\newglossary[slg]{symbolslist}{syi}{syg}{Symbols}
\newglossary[plg]{productlist}{pyi}{pyg}{Equipment, Materials, \& Software}
\newacronym{DiFC}{DiFC}{Diffuse \textit{in-vivo} Flow Cytometry}
\newacronym{CTC}{CTC}{Circulating Tumor Cell}

\newacronym{NIR}{NIR}{Near-InfraRed}
\newacronym{NIRS}{NIRS}{\gls{NIR} Spectroscopy}
\newacronym{CW}{CW}{Continuous-Wave}
\newacronym{TD}{TD}{Time-Domain}
\newacronym{SD}{\ensuremath{\text{SD}}}{Single-Distance}
\newacronym{SR}{\ensuremath{\text{SR}}}{Single-Ratio}
\newacronym{DR}{\ensuremath{\text{DR}}}{Dual-Ratio}
\newacronym{AF}{AF}{AutoFluorescence}
\newacronym{MC}{MC}{Monte-Carlo}
\newacronym{SNR}{SNR}{Signal-to-Noise Ratio}
\newacronym{PMT}{PMT}{PhotoMultiplier Tube}
\newacronym{PBS}{PBS}{Phosphate Buffered Saline}
\newacronym{FDA}{FDA}{Food and Drug Administration}

\newacronym{NADH}{NADH}{reduced Nicotinamide Adenine Dinucleotide}
\newacronym{FAD}{FAD}{Flavin Adenine Dinucleotide}

\newacronym[type=symbolslist]{dSD}{\ensuremath{\Delta\text{SD}}}{change in the \gls{SD}}
\newacronym[type=symbolslist]{dSR}{\ensuremath{\Delta\text{SR}}}{change in the \gls{SR}}
\newacronym[type=symbolslist]{dDR}{\ensuremath{\Delta\text{DR}}}{change in the \gls{DR}}
\newacronym[type=symbolslist]{rho}{\ensuremath{\rho}}{source-detector distance}
\newacronym[type=symbolslist]{r}{\ensuremath{\vec{r}}}{position vector}
\newacronym[type=symbolslist]{t}{\ensuremath{t}}{time}
\newacronym[type=symbolslist]{lam}{\ensuremath{\lambda}}{optical wavelength}
\newacronym[type=symbolslist]{PHI}{\ensuremath{\Phi}}{fluence rate}
\newacronym[type=symbolslist]{R}{\ensuremath{R}}{Reflectance}
\newacronym[type=symbolslist]{I}{\ensuremath{I}}{Intensity}
\newacronym[type=symbolslist]{Ir}{\ensuremath{\mathcal{I}}}{Intensity random variable}
\newacronym[type=symbolslist]{C}{\ensuremath{C}}{Coupling coefficient}
\newacronym[type=symbolslist]{Cr}{\ensuremath{\mathcal{C}}}{Coupling coefficient random variable}
\newacronym[type=symbolslist]{sigmaI}{\ensuremath{\sigma_{\acrshort{I}}}}{Intensity noise}
\newacronym[type=symbolslist]{sigmaRel}{\ensuremath{\sigma_{rel}}}{relative noise}
\newacronym[type=symbolslist]{mua}{\ensuremath{\mu_a}}{absorption coefficient}
\newacronym[type=symbolslist]{dmua}{\ensuremath{\Delta\mu_a}}{changes in \gls{mua}}
\newacronym[type=symbolslist]{mus}{\ensuremath{\mu_s}}{scattering coefficient}
\newacronym[type=symbolslist]{g}{\ensuremath{g}}{anisotropy factor}
\newacronym[type=symbolslist]{n}{\ensuremath{n}}{index of refraction}
\newacronym[type=symbolslist]{Q}{\ensuremath{Q}}{power volume density}
\newacronym[type=symbolslist]{P}{\ensuremath{P}}{Power}
\newacronym[type=symbolslist]{eta}{\ensuremath{\eta}}{fluorescence efficiency}
\newacronym[type=symbolslist]{V}{\ensuremath{V}}{Volume}
\newacronym[type=symbolslist]{W}{\ensuremath{W}}{fluorescence Jacobian}

\newcommand{\as}[1]{\acrshort{#1}}

\newcommand{\rl}[1]{#1}

\title{Dual-ratio approach for detection of point fluorophores in biological tissue}

\author[a,*,\textdagger]{Giles~Blaney}
\author[b,\textdagger]{Fernando~Ivich}
\author[a]{Angelo~Sassaroli}
\author[b]{Mark~Niedre}
\author[a]{Sergio~Fantini}
\affil[a]{Tufts University, Department of Biomedical Engineering, Medford, MA USA, 02155}
\affil[b]{Northeastern University, Department of Bioengineering, Boston, MA USA, 02120}

\cftpagenumbersoff{figure}
\cftpagenumbersoff{table} 
\begin{document} 
\maketitle

\begin{abstract}
    \\
    
    \noindent\textbf{Significance:} 
    Diffuse \textit{in-vivo} Flow Cytometry (DiFC) is an emerging fluorescence sensing method to non-invasively detect labeled circulating cells \textit{in-vivo}.
    However, due to Signal-to-Noise Ratio (SNR) constraints largely attributed to background tissue autofluorescence, DiFC's measurement depth is limited.
    \\
    multiplies
    \noindent\textbf{Aim:} 
    The Dual-Ratio (DR) / \rl{dual-slope} is a new optical measurement method that aims to suppress noise and enhance SNR to deep tissue regions.
    We aim to investigate the combination of DR and \rl{Near-InfraRed} (NIR) DiFC to improve circulating cells' maximum detectable depth and SNR.
    \\
    
    \noindent\textbf{Approach:} 
    Phantom experiments were used to estimate the key parameters in a diffuse fluorescence excitation and emission model.
    This model and parameters were implemented in Monte-Carlo to simulate DR DiFC while varying noise and autofluorescence parameters to identify the advantages and limitations of the proposed technique.
    \\
    
    \noindent\textbf{Results:} 
    Two key factors must be true to give DR DiFC an advantage over traditional DiFC; first, the fraction of noise that DR methods cannot cancel cannot be above the order of 10\% \rl{for acceptable SNR}.
    Second, DR DiFC has an advantage, \rl{in terms of SNR,} if the distribution of tissue autofluorescence contributors is surface-weighted.
    \\
    
    \noindent\textbf{Conclusions:} 
    DR cancelable noise may be designed for (\textit{e.g.} through the use of source multiplexing), and indications point to the autofluorescence contributors' distribution being truly surface-weighted \textit{in-vivo}.
    Successful and worthwhile implementation of DR DiFC depends on these considerations, but \rl{results} point to DR DiFC having possible advantages over traditional DiFC.
    
\end{abstract}


{\noindent \footnotesize\textbf{*}Giles~Blaney~Ph.D., \linkable{Giles.Blaney@tufts.edu} \\
\noindent \footnotesize\textbf{\textdagger} These authors contributed equally}


\section{Introduction}\label{sec:intro}
\Gls{DiFC} is an emerging optical technique that enables fluorescence detection of rare circulating cells in the bloodstream in the optically diffusive medium\cite{tan_vivoflowcytometry_2019, patil_fluorescencemonitoringrare_2019, pace_nearinfrareddiffusevivo_2022, zettergren_instrumentfluorescencesensing_2012}.
\rl{dual-slope} or \gls{DR} is a new diffuse optical technique that is designed to suppress noise in the optical signal and reduce sensitivity to superficial tissue regions \cite{sassaroli_dualslopemethodenhanced_2019, fantini_transformationalchangefield_2019, blaney_phasedualslopesfrequencydomain_2020, blaney_methodmeasuringabsolute_2022}.
A challenge of \gls{DiFC} is the contamination of the target fluorescence signal from noise, which may be associated with background \gls{AF}. 
In this work, we investigate the possibility of utilizing \gls{DR} techniques to suppress this noise and \gls{AF}, thus enabling better \gls{SNR} of \gls{DiFC} measurements.
\par

\subsection{Diffuse \textit{in-vivo} Flow Cytometry}\label{sec:intro:difc}
In \gls{DiFC}, the tissue surface is illuminated with laser light, typically delivered by an optical fiber bundle.
As fluorescently-labeled circulating cells pass through the \gls{DiFC} field of view, a transient fluorescent peak may be detected at the surface using a collection fiber.
Since \gls{DiFC} uses diffuse light, it is possible to detect circulating cells from relatively deep-seated and large blood vessels several \si{\milli\meter} into tissue. 
Hence, the key advantage of \gls{DiFC} is that it allows for the interrogation of relatively large circulating blood volumes enabling the detection of rare cells. 
\par

For example, a major application of \gls{DiFC} has been in mouse cancer research.
\Glspl{CTC} migrate from solid tumors, move through the circulatory system, and may form secondary tumor sites.
As such, \glspl{CTC} are instrumental in hematogenous cancer metastasis and are a major focus of medical research. 
However, \glspl{CTC} are extremely rare and frequently number fewer than \SI{100}{\acrshort{CTC}\per\milli\liter} of blood in metastatic patients\cite{pang_circulatingtumourcells_2021}.
\rl{In the clinic, the current gold standard method to study \glspl{CTC} is via liquid biopsy (blood draws) using the \gls{FDA} cleared system CellSearch.\cite{Alix-Panabieres_ClinicalChemistry13_CirculatingTumor,Mader_OncologyResearchandTreatment17_LiquidBiopsy}
For example, in breast cancer patients, $\ge$\SI{5}{\acrshortpl{CTC}} detected in \SI{7.5}{\milli\liter} blood samples with CellSearch is associated with poor cancer prognosis.\cite{pang_circulatingtumourcells_2021}
However, the small amount of blood analyzed (\SI{0.015}{\percent} of the total peripheral blood volume) is known to yield poor sampling statistics, even assuming ideal Poisson statistics.\cite{Mishra_Proc.Natl.Acad.Sci.20_UltrahighthroughputMagnetic}
Moreover, short-term temporal fluctuations of these cells may cause errors in estimating the \gls{CTC} numbers based on when blood was drawn.\cite{Diamantopoulou_Nature22_MetastaticSpread}
Thus, \gls{DiFC} to enumerate \glspl{CTC} continuously and non-invasively in large circulating blood volumes \textit{in-vivo} may help address these limitations.}
In mouse xenograft tumor models, we have previously performed \gls{DiFC} on the mouse tail above the ventral caudal artery.
These studies showed that it is possible to non-invasively sample the entire peripheral blood volume in approximately \SI{15}{\minute}, permitting the detection of very rare \glspl{CTC}.\cite{patil_fluorescencemonitoringrare_2019, williams_shorttermcirculatingtumor_2020}
\par

\begin{figure}
    \begin{center}
        \includegraphics[width=0.9\linewidth]{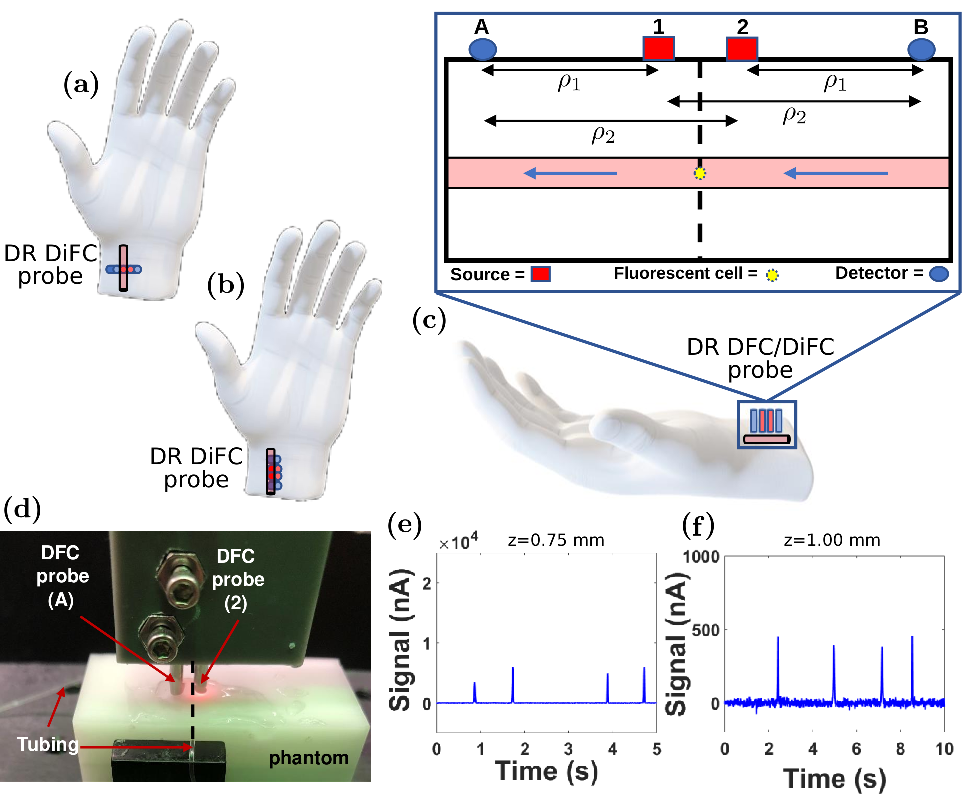}
    \end{center}
    \caption{\label{fig:introScheme} 
    Conceptual application of \acrfull{DR} \acrfull{DiFC}. In principle, source and detector pairs could be arranged \textbf{(a)} perpendicular, or \textbf{(b)} parallel to the underlying artery (in this case, the radial artery). \textbf{(c)} Use of two sources (\texttt{1}, \texttt{2}) and two detectors (\texttt{A}, \texttt{B}) would permit four source and detector pairs separated by a \acrfull{rho}. \textbf{(d)} Photograph of \acrfull{NIR} \acrfull{DiFC} system\cite{pace_nearinfrareddiffusevivo_2022} on a diffusive flow phantom with fiber probes arranged perpendicular to the tubing direction. \rl{The dashed black line in \textbf{(d)} corresponds to the mid-plane shown in \textbf{(c)}.} The instrument permitted measurement with a single source and detector pair, in this case (\texttt{2} \& \texttt{A}). Background-subtracted sample \acrshort{DiFC} data of fluorescent microspheres embedded \textbf{(e)} \SI{0.75}{\milli\meter} deep and \textbf{(f)} \SI{1.00}{\milli\meter} deep in a phantom with a flow channel.
    }
\end{figure}

Moreover, if paired with specific molecular contrast agents for \glspl{CTC} \rl{(\textit{e.g.}, as those for fluorescence-guided surgery)}, translation of \gls{DiFC} to humans may be possible\cite{lee_reviewclinicaltrials_2019, hernot_latestdevelopmentsmolecular_2019, niedre_prospectsfluorescencemolecular_2022}.
We recently demonstrated that \glspl{CTC} could be labeled directly in the blood of mice by using a folate-targeted fluorescent target (EC-17) and be detected externally with \gls{DiFC}\cite{patil_fluorescencelabelingcirculating_2020}.
However, two major technical challenges exist when applying \gls{DiFC} to humans.
First, the measured \gls{DiFC} signal combines a fluorescence signal from the circulating cell and a non-specific background \gls{AF} signal from the surrounding tissue.
Although the non-specific background signal can be subtracted, noise contributed from the background cannot be removed.
As such, the noise may obscure small-amplitude fluorescence signals.
To illustrate this \gls{SNR} consideration, example \gls{DiFC} data showing signal peaks from detected flowing fluorescent microspheres embedded \SI{0.75}{\milli\meter} deep and \SI{1.00}{\milli\meter} deep in a phantom are shown in Figure~\ref{fig:introScheme}\textbf{(e)(f)}, respectively.
\rl{Note that each peak represents individual mirco-spheres passing through \gls{DiFC} field-of-view.}
The second technical challenge relates to the depth of blood vessels in humans.
Suitable blood vessels such as the radial artery in the wrist (Figure~\ref{fig:introScheme}\textbf{(a)-(c)}) are expected to be about \SIrange{2}{4}{\milli\meter} deep (\textit{i.e.}, significantly deeper than in mice)\cite{selvaraj_ultrasoundevaluationeffect_2016}.
\rl{Alignment to the radial artery can be potentially achieved in humans by placing \gls{DiFC} probes in the \textit{volar} wrist beneath the thumb, between the wrist bone and the tendon.
Some people may be able to observe the vessel through the skin, facilitating alignment.
Small misalignments (approximately \SI{1}{\milli\meter}) do not significantly affect the signal quality because of the wide sensitivity volume of diffuse light.}
We recently showed that individual cell fluorescence signals might be detectable up to \SI{4}{\milli\meter} deep using \gls{NIR} fluorophores and a \gls{rho} of approximately \SI{3}{\milli\meter}, increased depth and larger relative non-specific background signals presents challenges for detection of weakly-labeled \glspl{CTC}\cite{ivich_signalmeasurementconsiderations_2022}. Hence methodology for reduction of the background signal and its contributed noise, such as \rl{dual-slope} or \gls{DR} described in Section~\ref{sec:intro:ds}, would be extremely valuable for potential human translation of \gls{DiFC}. 
\par

\subsection{Dual-Ratio}\label{sec:intro:ds}
One of the most basic measurements in diffuse optics is that of the \gls{SD} \gls{R}, which is measured by placing a point source and a point detector on the surface of a highly scattering medium.
In practice, the detector does not measure the theoretical \gls{R} but instead, some optical \gls{I} that is proportional to the theoretical \gls{R} (Appendix~\ref{sec:app:noise:coup}).
The proportionality of \gls{I} to \gls{R} is controlled by \glspl{C}, which may be associated with the coupling efficiency of the optodes (sources or detectors) with the medium as well as any other multiplicative factor on \gls{R}.
\par

Recently, we introduced new \rl{measurement} types that attempt to represent the theoretical \gls{R} by canceling any \glspl{C} associated with optodes.
These are the \rl{dual-slope}\cite{blaney_phasedualslopesfrequencydomain_2020} or the \gls{DR}\cite{blaney_methodmeasuringabsolute_2022} data types, which in their calculation cancel any multiplicative factors associated with optodes\cite{blaney_functionalbrainmapping_2022, blaney_methodmeasuringabsolute_2022}.
These measurement types essentially recover the spatial dependence of \gls{R} over multiple \glspl{rho}.
This is achieved from a symmetrical optode arrangement of sources and detectors such as the one in Figure~\ref{fig:introScheme}\textbf{(c)}.
\par

A further advantage of the \rl{dual-slope} or \gls{DR} data type is the reduced contribution to the signal from superficial parts of the diffuse medium\cite{sassaroli_dualslopemethodenhanced_2019, fantini_transformationalchangefield_2019, blaney_phasedualslopesfrequencydomain_2020}.
With the hypothesis that the unwanted \gls{AF} contributors in \gls{DiFC} measurements are mainly near the surface, we decided to explore if these data types could suppress the \gls{AF} component of the \gls{DiFC} measurement.
We hypothesize that this \gls{AF} suppression, together with the cancellation of noise through the cancellation of \glspl{C}, will improve the \gls{SNR} of \gls{DR} over traditional \gls{SD} \gls{DiFC} measurements using \gls{NIR} \glspl{lam}.
This improved \gls{SNR} of \gls{DR} \gls{DiFC} could enable the detection of deep fluorescent targets, which is a current struggle of \gls{DiFC} as discussed in Section~\ref{sec:intro:difc}.
Therefore, this work investigates the potential of \gls{DR} in the \gls{DiFC} application and identifies the key parameters which control whether or not \gls{DR} will be advantageous.
\par

\section{Methods}\label{sec:meth}
\subsection{Proposed Dual-Ratio Signal}\label{sec:meth:sig}
Most diffuse optical measurements recover an optical signal using a source and a detector, often each placed on a tissue surface. 
We consider each measurement of \gls{I} with a source and detector as a \gls{SD} measurement (Appendix~\ref{sec:app:defs:SD}).
Previous \gls{DiFC} work has considered these types of measurements\cite{ivich_signalmeasurementconsiderations_2022}; however, in this work, we introduce a \gls{DiFC} measurement type based on a combination of \glspl{SD} to form a \gls{DR} (Appendix~\ref{sec:app:defs:DR})\cite{blaney_phasedualslopesfrequencydomain_2020, blaney_methodmeasuringabsolute_2022}.
\par

\Gls{DR} is defined in detail in Appendix~\ref{sec:app:defs:DR}, and will be summarized here.
The \gls{DR} measurement is defined as follows:
\begin{equation}\label{equ:DR}
    \acrshort{DR}=\sqrt{\frac{\acrshort{I}_{l,\text{\texttt{I}}}\acrshort{I}_{l,\text{\texttt{II}}}}{\acrshort{I}_{s,\text{\texttt{I}}}\acrshort{I}_{s,\text{\texttt{II}}}}}
\end{equation}
\noindent where the $l$ and $s$ subscripts represent \gls{I} measurements at long or short \glspl{rho}, respectively.
Additionally, the \texttt{I} and \texttt{II} subscripts show whether it is the first or second symmetric \gls{I} measurement in a \gls{DR}. 
The exact geometric requirements for \texttt{I} and \texttt{II} are described in more detail in Appendix~\ref{sec:app:defs:DR} and previous publications (where \gls{DR} is referred to as \rl{dual-slope, since geometric requirements for the two are the same})\cite{blaney_designsourcedetector_2020}. 
For this work, consider Figure~\ref{fig:introScheme}(\textbf{c}).
In this case, $s$,\texttt{I} and $l$,\texttt{I} are detector \texttt{A} \& source \texttt{1} (\gls{SD} \texttt{A1}) and \gls{SD} \texttt{A2}, respectively.
And similarly, for detector \texttt{B}, $s$,\texttt{II} and $l$,\texttt{II} are \gls{SD} \texttt{B2} and \gls{SD} \texttt{B1}, respectively.
\par

\subsection{Monte-Carlo Models}\label{sec:meth:mc}
This work presents simulation results based on a \gls{MC} model run in Monte-Carlo~Extreme \rl{(MCXLAB~2020 running in MATLAB~2023a on an NVIDIA~RTX~4090)}\cite{fang_montecarlosimulation_2009}.
For all simulations optical properties meant to represent \gls{lam} of \SI{810}{\nano\meter} were used\cite{ivich_signalmeasurementconsiderations_2022}.
These were an \gls{mua} of \SI{0.002}{\per\milli\meter}, a \gls{mus} of \SI{7}{\per\milli\meter}, an \gls{g} of \num{0.9}, and an \gls{n} of \num{1.37}.
The \gls{MC} was run by launching \num{e9} photons into a $\SI{30}{\milli\meter}\times\SI{30}{\milli\meter}\times\SI{30}{\milli\meter}$ volume, with grid spacing of \SI{0.1}{\milli\meter} and \num{10} time gates ending at \SI{10}{\nano\second}.
To calculate the \gls{CW} \gls{R}, the \gls{MC}'s \gls{TD} \gls{R} was integrated over the time gates.
\par

For the coordinate system, the surface of the medium was considered to be $z=\SI{0}{\milli\meter}$ (positive $z$ pointing into the medium), and the center (with the edges \SI{15}{\milli\meter} away in each direction) to be $x=\SI{0}{\milli\meter}$ and $y=\SI{0}{\milli\meter}$.
Using this coordinate system, detector \texttt{A} was placed at $-3.5\hat{x}~\si{\milli\meter}$, detector \texttt{B} at $3.5\hat{x}~\si{\milli\meter}$, source \texttt{1} at $-0.5\hat{x}~\si{\milli\meter}$, and source \texttt{2} at $0.5\hat{x}~\si{\milli\meter}$.
In the case of the \SI{0}{\milli\meter} \gls{rho} simulation, both source and detector were placed at the origin.
A separate \gls{MC} was run for each of these optodes to find the \gls{PHI} distribution for a \gls{MC} source placed at the optode position.
This follows the adjoint method to calculate the \gls{R} for a given \gls{SD} set\cite{Yao_Biomed.Opt.ExpressBOE18_DirectApproach}.
In the case of sources, a \gls{MC} pencil beam was used, and in the case of detectors, a \gls{MC} cone beam with a numerical aperture of \num{0.5} was used to simulate the collection geometry of a fiber more accurately.
This type of source distribution associated with the detectors is done to be closer to the condition of validity of the reciprocity theorem, which is crucial for applying the adjoint method.
These \glspl{PHI} were used with Equation~\ref{equ:app:W} to yield the \gls{W}\cite{ivich_signalmeasurementconsiderations_2022}, a key parameter used in all the simulation results as described in Appendix~\ref{sec:app:theo}.
\par

\subsection{Experimental Measurements with Diffuse Flow Cytometry}\label{sec:meth:exp}
\subsubsection{Phantom Experiments with Fluorescent Microspheres \textit{In-Vitro}}\label{sec:meth:exp:phantom}
To estimate the potential performance of a \gls{DR} \gls{DiFC} instrument, we performed \gls{DiFC} measurements in tissue-mimicking flow phantoms with \gls{NIR} wavelengths \rl{(\SI{770}{\nano\meter} excitation and \SI{810}{\nano\meter} emission)}\cite{ivich_signalmeasurementconsiderations_2022}. 
Briefly, we used Jade Green High Intensity (JGHI, Spherotech Inc., Lake Forest, IL USA) cell-mimicking fluorescent microspheres (size approximately \SIrange{10}{14}{\micro\meter}) for \gls{NIR} \gls{SD} measurements\cite{pace_nearinfrareddiffusevivo_2022}.
Microspheres were suspended at a concentration of \SI{e3}{\per\milli\liter} in \gls{PBS} and flowed through Tygon tubing (internal diameter of \SI{0.25}{\milli\meter}; TGY-010-C, Small Parts Inc., Seattle, WA USA) embedded in a tissue-mimicking optical flow phantom made of high-density polyethylene.
Microsphere suspensions were pumped with a syringe pump (0-2209, Harvard Apparatus, Holliston, MA USA) at a flow rate of \SI{50}{\micro\liter\per\minute}, or average flow velocity of \SI{17}{\milli\meter\per\second} \rl{to approximate linear velocities in the mouse leg femoral artery}.\cite{patil_fluorescencemonitoringrare_2019, Hartmann_Phys.Med.Biol.17_FluorescenceDetection}
The tubing at a depth of \SI{1.50}{\milli\meter} in the phantom.
\par

We performed \gls{SD} \gls{DiFC} measurements using a \gls{rho} of \SI{3}{\milli\meter} (Figure~\ref{fig:introScheme}\textbf{(c)} $\acrshort{rho}_1$).
Two fiber bundle probes were used, one as a source fiber and the other as a detector fiber.
We collected \gls{DiFC} data for the source (labeled as numbers) and detector (labeled as letters) pairs \texttt{A1} and \texttt{B2}.
To test the effect of the geometric arrangement of the probes relative to the flow tube, we performed measurements in the perpendicular or parallel directions as in Figure~\ref{fig:introScheme}\textbf{(a)(b)}.
For illustration, a photograph of the \texttt{A2} (at $\acrshort{rho}_2=\SI{4}{\milli\meter}$) \gls{SD} \gls{NIR} \gls{DiFC} measurement is included in Figure~\ref{fig:introScheme}\textbf{(d)}, which shows a perpendicular placement of the probes to the tubing.
\par

We collected \SI{15}{\minute} \gls{DiFC} scans and processed the data as described previously\cite{tan_vivoflowcytometry_2019}.
First, we performed background subtraction using a \SI{5}{\second} moving median window.
Transit fluorescent peaks were measured as fluorescent microspheres were detected with the \gls{DiFC} system.
Here, peaks were defined as having a minimum amplitude of \rl{five} times the noise (standard deviation of the data in a \SI{1}{\minute} moving window).
We analyzed peak amplitude, width (\textit{i.e.}, shape), and noise properties. 
These parameters were used in Appendix~\ref{sec:app:expPar} to find the important parameters of \gls{eta} and fluorescent \gls{mua} needed to simulate \gls{SNR} of \gls{DiFC} measurements from the \gls{MC} output.
\par

\subsubsection{Measurement of \textit{In-Vivo} Autofluorescence on Mice}\label{sec:meth:exp:miceaf}
As discussed in more detail in Section~\ref{sec:dis:depth} and Appendix~\ref{sec:app:expPar:bck}, knowledge of the distribution of sources of \gls{AF} is necessary for building a computational model of \gls{DR} \gls{DiFC} noise and background.
To address this, we measured the \gls{AF} background \gls{DiFC} signal of the shaved inner thigh of recently euthanized mice: (1) at the surface of the skin and (2) at the surface of the exposed underlying muscle (without skin).
\par

We used $N=\num{4}$ \rl{female} BALB/c mice \rl{on a low-fluorescence chow diet} for this experiment.
The hair of the inner left thigh of these euthanized mice was removed with depilatory cream (Nair, Church \& Dwight Co., Inc., Ewing, NJ).
The skin of the right inner thigh was surgically removed, exposing the underlying muscle.
We used our red \gls{DiFC} instrument\cite{tan_vivoflowcytometry_2019} to measure the background tissue \gls{AF} background \gls{SD} signal in both cases.
During these measurements, ultrasound gel was applied between the \gls{DiFC} probe and the tissue surface to minimize the index of refraction mismatch.
\gls{DiFC} was performed for \SI{15}{\minute}, and the mean background (\textit{i.e.}, with and without skin) was calculated from the resulting data.
\par

\section{Results}\label{sec:res}
\subsection{Monte-Carlo Simulated Maps \& Signals}\label{sec:res:mcsim}
\begin{figure}[pth]
    \begin{center}
        \includegraphics{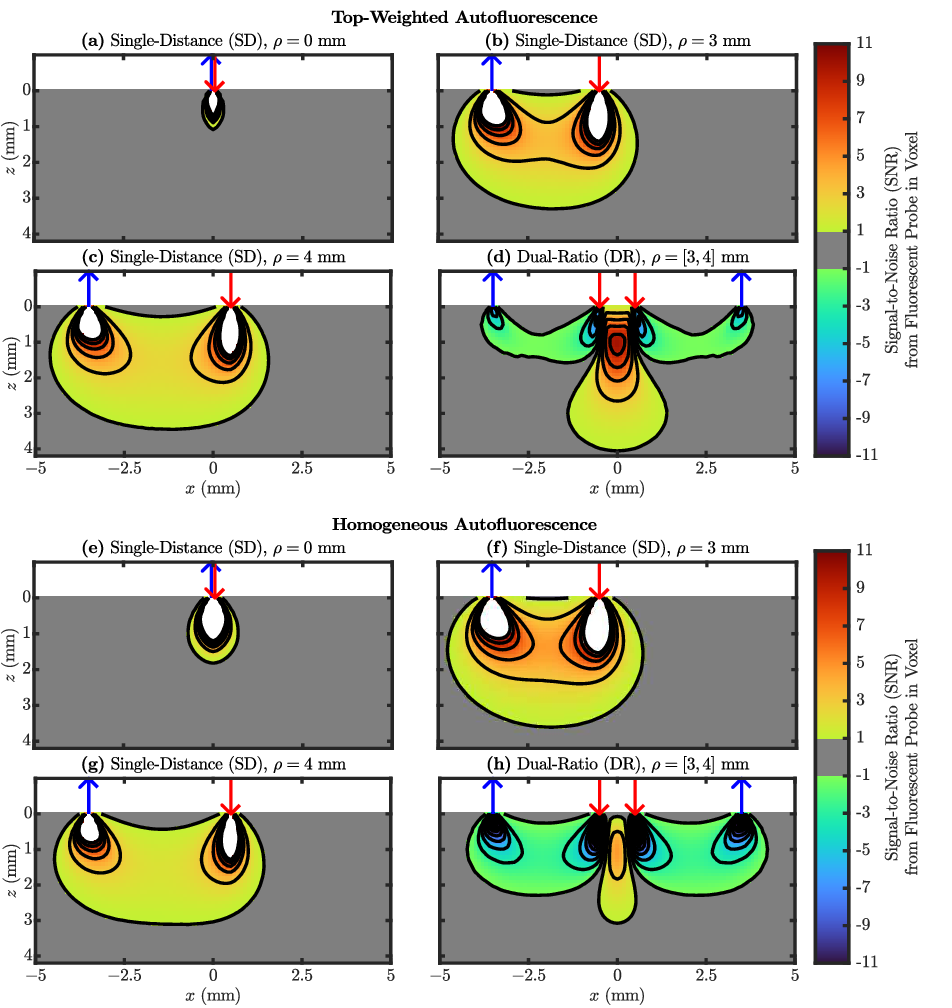}
    \end{center}
    \caption 
    { \label{fig:SNRmap}
    Map of the \acrfull{SNR} to a fluorescent target at a particular position (in the $y=\SI{0}{\milli\meter}$ plane) within a medium with surface-weighted (\textbf{(a)}-\textbf{(d)}) or homogeneous (\textbf{(e)}-\textbf{(h)}) \acrfull{AF} contributors for four different measurement types with detectors represented by blue arrows and sources by red arrows.
    \textbf{(a),(e)} \Acrfull{SD} at a \acrfull{rho} of \SI{0}{\milli\meter}.
    \textbf{(b),(f)} \acrshort{SD} at a \acrshort{rho} of \SI{3}{\milli\meter}.
    \textbf{(c),(g)} \as{SD} at a \as{rho} of \SI{4}{\milli\meter}.
    \textbf{(d),(h)} \Acrfull{DR} containing \acrshortpl{rho} of \SIlist{3;4}{\milli\meter}. \\
    Note: White regions represent \as{SNR} greater than the maximum color-bar scale \rl{(\textit{i.e.}, 11)} and gray regions represent absolute \as{SNR} less than one.\\
    Parameters: Source~=~pencil, Detector~=~\SI{0.5}{NA} cone, voxel~=~$\SI{0.1}{\milli\meter}\times\SI{0.1}{\milli\meter}\times\SI{0.1}{\milli\meter}$, \acrfull{mua}~=~\SI{0.002}{\per\milli\meter}, \acrfull{mus}~=~\SI{7}{\per\milli\meter}, \acrfull{g}~=~\num{0.9}, \acrfull{n}~=~\num{1.37}, for surface-weighted (\textbf{(a)}-\textbf{(d)}) \as{AF} \acrfull{eta}~$\propto e^{\ln{(0.5)}z/\SI{0.1}{\milli\meter}}$ for homogeneous (\textbf{(e)}-\textbf{(h)}) \as{eta} constant, \& signal and noise parameters found in appendix~Appendix~\ref{sec:app:expPar} (in this simulation we assumed \SI{5}{\percent} Non-Cancelable noise).
    }
\end{figure}

In Figure~\ref{fig:SNRmap}, we show the simulated \gls{SNR} from a fluorescent target placed anywhere on the $y=\SI{0}{\milli\meter}$ plane. 
This shows which possible target positions would yield a measurable signal (\gls{SNR} greater than one) and how strong the signal would be.
One striking result that may be drawn from comparing the overall extent of the region with \gls{SNR} greater than one for the case with surface-weighted contributions of \gls{AF} (Figure~\ref{fig:SNRmap}\textbf{(a)}-\textbf{(d)}) and the case with homogeneous contributions (Figure~\ref{fig:SNRmap}\textbf{(e)}-\textbf{(h)}).
In general, the case of homogeneous contributions of \gls{AF} favors the shorter \glspl{rho}. 
It also severely weakens the ability of \gls{DR} to measure targets when they are deep within the medium.
In-fact, close examination of Figure~\ref{fig:SNRmap}\textbf{(f)}\&\textbf{(g)} suggests that, counter-intuitively, a shorter \gls{rho} can measure deeper.
However, this is only evident in the homogeneous \gls{AF} contributors case and not in the surface-weighted case.
Thus telling us that the advantages of one measurement type over another depend on the spatial distribution of \gls{AF} contributors.
\par

We may also compare the different \glspl{rho} and the \gls{SD} versus \gls{DR}. 
For a \SI{0}{\milli\meter} \gls{rho}, we see a small bulb shape close to the surface, confirming what is expected for a co-localized source and detector.
The non-zero \glspl{rho} for \gls{SD} show the typical banana shape expected in diffuse optics but with some differences.
These differences arise from the small length scale used in these simulations, on the order of \SI{1}{\milli\meter} so that the light does not act fully diffuse.
For this reason, the \gls{MC} source model matters much for the shape of the banana.
To be more realistic, we modeled the source as a pencil beam and the detector as a cone with \num{0.5} numerical aperture. 
This resulted in one side of the banana (the one near the source) being deeper than the other since the pencil beam could more effectively launch photons along the $z$ direction into the medium. 
These observations show that the \gls{MC} source condition simulated matters and should match reality.
\par

Focusing on \gls{DR}, we see that it can measure deeper than every other measurement in the surface-weighted \gls{AF} contributors case.
However, in the homogeneous case, it does no better than the long (\gls{rho} of \SI{4}{\milli\meter}) \gls{SD}.
This is primarily because \gls{DR} focuses on canceling out signals from close to the surface.
Therefore, it is advantageous when the confounding signal's contributors, the \gls{AF}'s chromophore concentration or efficiency, is surface weighted.
This further reinforces the importance of understanding the distribution of \gls{AF} in a realistic measurement case, like tissue, since it heavily affects which measurement type is preferable when detecting deep fluorescent targets.
\par

\begin{figure}[pth]
    \begin{center}
        \includegraphics{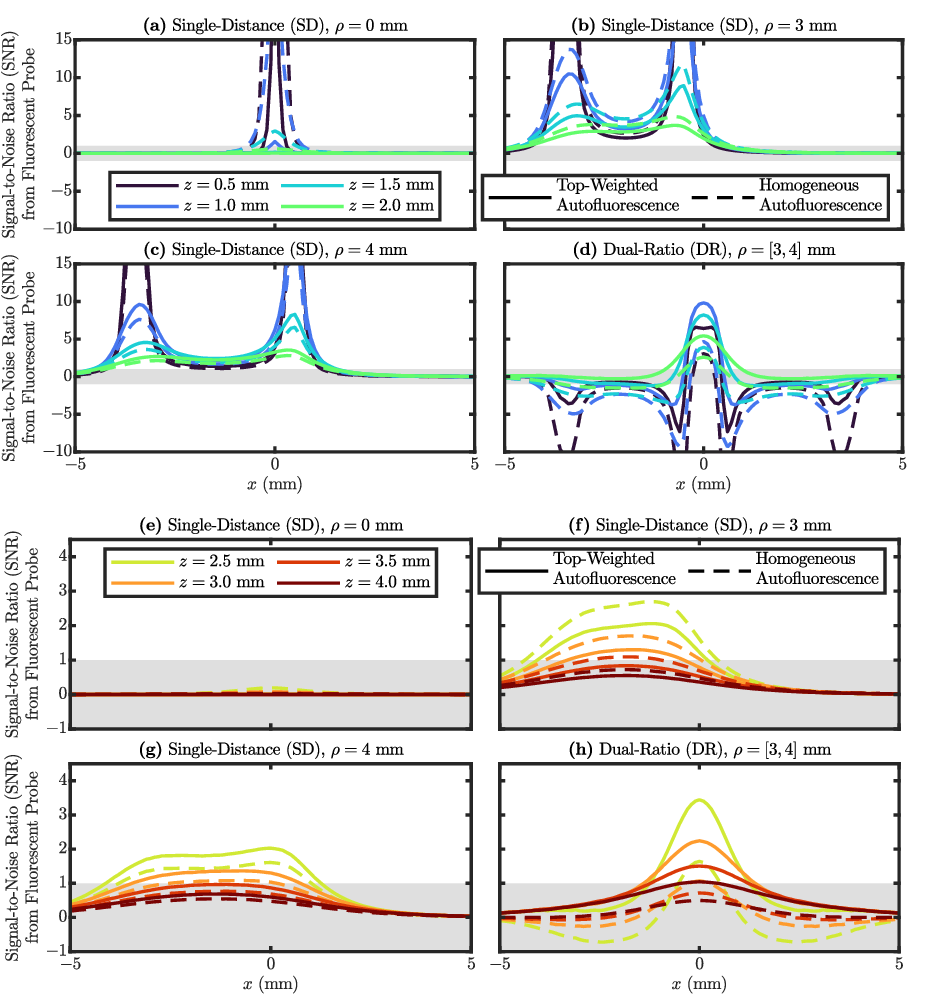}
    \end{center}
    \caption 
    { \label{fig:SIGsim}
    Traces of expected \acrfull{SNR} from a fluorescent target flowing at a particular depth (color) beneath the source-detector arrangement (same as Figure~\ref{fig:introScheme},\ref{fig:SNRmap}; $y=\SI{0}{\milli\meter}$).
    These results are shown for surface-weighted and homogeneous \gls{AF} contributors (line-type).
    \textbf{(a),(e)} \Acrfull{SD} at a \acrfull{rho} of \SI{0}{\milli\meter}.
    \textbf{(b),(f)} \acrshort{SD} at a \acrshort{rho} of \SI{3}{\milli\meter}.
    \textbf{(c),(g)} \as{SD} at a \as{rho} of \SI{4}{\milli\meter}.
    \textbf{(d),(h)} \Acrfull{DR} containing \acrshortpl{rho} of \SIlist{3;4}{\milli\meter}. \\
    Note: Gray regions represent absolute \as{SNR} less than one. \\
    Parameters: The same as Figure~\ref{fig:SNRmap} with assumed \SI{5}{\percent} Non-Cancelable noise. 
    } 
\end{figure}

In Figure~\ref{fig:SIGsim}, we present the expected signal profiles (in \gls{SNR}) that one would measure if a fluorescent target flowed parallel to the source-detector line (in the $y=\SI{0}{\milli\meter}$ plane).
The x-axis shows the $x$ coordinate; in an actual \gls{DiFC} measurement, this would be time with the $x$ position scaled by the velocity.
Different colors represent targets flowing at different depths, and line type shows the type of distribution of contributors to \gls{AF}.
The gray region shows where the absolute \gls{SNR} is less than one, so if the curve drops into this region, we may say it is not measurable.
\par

Many of the conclusions drawn from Figure~\ref{fig:SIGsim} are similar to the ones that one may draw from Figure~\ref{fig:SNRmap} since these traces (Figure~\ref{fig:SIGsim}) are simply line scans of Figure~\ref{fig:SNRmap}.
So, we focus on more apparent features in these traces than the map.
The first is the shape of a non-zero \gls{rho} \gls{SD} measurement.
This is that of a double peak, which arises when the target first flows under the detector, making a strong signal, then under the source causing the target to make another strong signal.
This is especially evident with shallow depths but is slightly present even in the broad traces simulated from deep $z$s.
Further, because of the different \gls{MC} source conditions, the peak height when the target passes under the source is higher than when the target passes under the detector.
Therefore, we may say that this model predicts that a non-zero \gls{rho} \gls{SD} measurement will produce a rather unique signal, with a double peak and a more significant peak height when the target passes under the source.
\par

Finally, lets examine the shape of the simulated \gls{DR} signal (Figure~\ref{fig:SIGsim}\textbf{(d)},\textbf{(h)}).
In this case, the signal has positive and negative components. 
Here it is essential to understand that these traces are differences from a baseline measurement, so that Figure~\ref{fig:SIGsim}\textbf{(d)},\textbf{(h)} shows the baseline \gls{DR} subtracted from the current \gls{DR}.
This is detailed in Appendix~\ref{sec:app:defs} and Equation~\ref{equ:app:dDR}.
Therefore negative values in the \gls{DR} signal mean that the current \gls{DR} value is less than the baseline value.
Since one wishes to identify the presence of a fluorescent target, it does not matter whether the signal is positive or negative, so even \gls{SNR} of \gls{DR} less than \num{-1} may be considered a signal which can identify the target.
Knowing this, we see that the \gls{DR} signal is unique and could further help identify if a target is genuinely detected.
The \gls{DR} signal is expected first to decrease as the target flows under the detector, then an increase when the target is under the source, followed by a decrease when the target finally flows under the second detector.
Therefore, these results suggest that the \gls{DR} (as well as the \gls{SD}) signal has a unique shape that may be used to identify a true positive detection of the fluorescent target.
\par

\subsection{Comparison of Phantom Experiments and Simulated Signals}\label{sec:res:exmccomp}
\begin{figure}[pth]
    \begin{center}
        \includegraphics{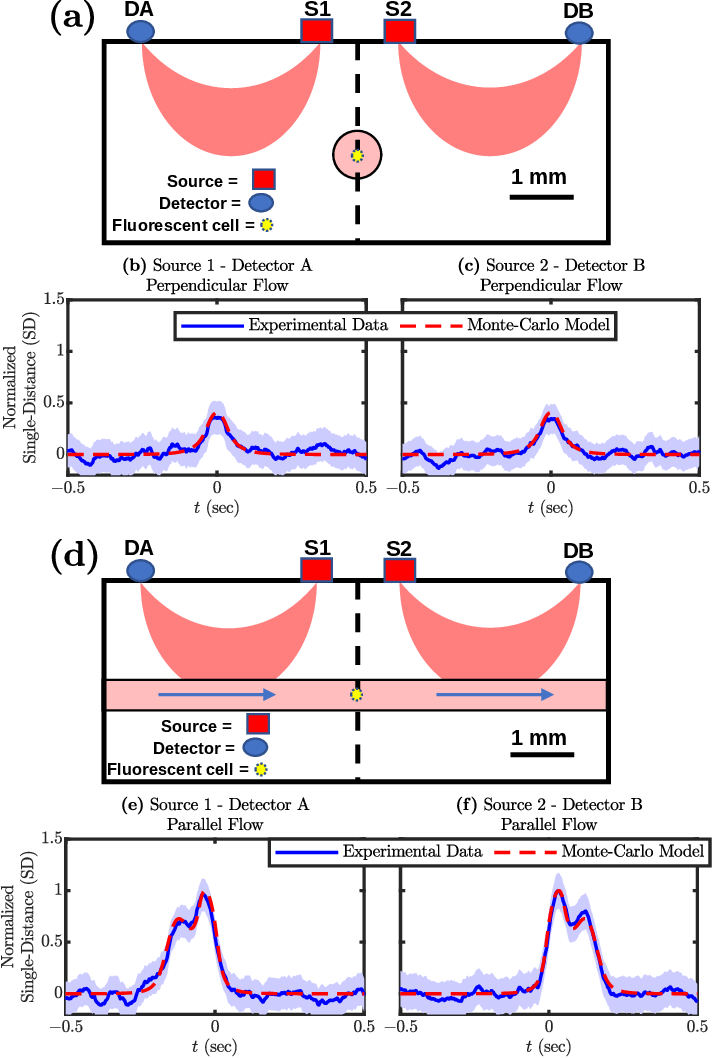}
    \end{center}
    \caption 
    { \label{fig:expVsim}
    Traces showing a comparison of experimental phantom data and expected results from the \acrfull{MC} model. \Acrfull{SD} traces are normalized so that the mean peak maximum between panels \textbf{(e)} and \textbf{(f)} is one, meaning all sub-panels utilize the same normalization factors. \textbf{(a)} Schematic of perpendicular flow case. \textbf{(b),(c)} Comparison for perpendicular flow case. \textbf{(d)} Schematic of parallel flow case. \textbf{(e),(f)} Comparison for parallel flow case.\\
    Note: Shaded regions represent the noise level of the experimental data. \\
    Parameters: The same as Figure~\ref{fig:SNRmap} with a known of \SI{1.5}{\milli\meter} and assumed fluorescent target velocity of \SI{25}{\milli\meter\per\second}. 
    } 
\end{figure}

We collected experimental \gls{DiFC} data using a phantom and fluorescent microspheres as described in Section~\ref{sec:meth:exp:phantom}. 
Figure~\ref{fig:expVsim} shows sample normalized and smoothed (with the shaded region showing noise) fluorescent microsphere peaks (\textit{i.e.}, detections) flowing \SI{1.50}{\milli\meter} deep in a phantom.
\gls{DiFC} measurements from two \gls{SD} pairs, \texttt{A1} and \texttt{B2}, are shown in perpendicular (Figure~\ref{fig:expVsim}\textbf{(a)-(c)}) and parallel (Figure~\ref{fig:expVsim}\textbf{(d)-(f)}) probe to tube configurations.
Normalization was performed to set the mean peak maximum of Figure~\ref{fig:expVsim}\textbf{(e)}\&\textbf{(f)} to one while using the same normalization factor for Figure~\ref{fig:expVsim}\textbf{(b)}\&\textbf{(c)}.
This allows for comparing the relative amplitudes between all traces in Figure~\ref{fig:expVsim}.
\par

So that one may relate these normalized values to measured \gls{PMT} currents, we provide the experimentally measured amplitudes for this data.
For the parallel tube to probe configuration (Figure~\ref{fig:expVsim}\textbf{(e)}-\textbf{(f)}), \gls{DiFC} peak maximum amplitudes averaged \SI{41}{\nano\ampere} and \SI{38}{\nano\ampere} for \texttt{A1} and \texttt{B2}, respectively.
Meanwhile, for the perpendicular tube to probe configuration (Figure~\ref{fig:expVsim}\textbf{(a)}\&\textbf{(c)}), the peak amplitudes averaged \SI{16.64}{\nano\ampere} and \SI{17.22}{\nano\ampere} for \texttt{1A} and \texttt{2B}, respectively. 
\par

We have also overlaid normalized peaks simulated using \gls{MC} (red dashed lines), using the same normalization method as the experimental data.
This type of normalization ensures the mean amplitudes match experimental data for Figure~\ref{fig:expVsim}\textbf{(e)}\&\textbf{(f)} but does not ensure a match for Figure~\ref{fig:expVsim}\textbf{(b)}\&\textbf{(c)} or the individual amplitudes Figure~\ref{fig:expVsim}\textbf{(e)}\&\textbf{(f)}.
Additionally, the simulated velocity for the \gls{MC} data was chosen to match the peak width with the experimental data.
For this, a velocity of \SI{25}{\milli\meter\per\second} was used.
However, since this velocity was assumed for all panels of Figure~\ref{fig:expVsim}, a match to the peak width for an individual panel is not ensured.
Therefore, a comparison between experimental data and \gls{MC} simulations should be done against features that were not made to match.
These features were the amplitude of the peaks in Figure~\ref{fig:expVsim}\textbf{(b)}\&\textbf{(c)}, the amplitude of the lesser second peak in Figure~\ref{fig:expVsim}\textbf{(e)}\&\textbf{(f)}, and the overall shape of the signal.
With this in mind, we see excellent agreement between the \gls{MC} simulations and experimental measurements.
This is particularly true regarding the doublet peak shown in Figure~\ref{fig:expVsim}\textbf{(e)}\&\textbf{(f)}, which is predicted by \gls{MC} and observed experimentally.
Note that the double peak in the parallel configuration is generated by the microsphere passing through high values of \gls{W} twice, once under the detector and once under the source.
Additionally, the relative amplitudes of the doublet are influenced by the relationship between the numerical aperture of the source and detector, reinforcing the choice of a cone instead of a pencil beam for the detector in the \gls{MC} simulation.
In fact, if a pencil beam was chosen for both the source and detector \glspl{MC}, the match to experimental data would be poor as the doublet peak would have the same maximum for both peaks in the doublet. 
\par

\section{Discussion}\label{sec:dis}
\subsection{Expected Depth}\label{sec:dis:depth}
\begin{figure}[htb]
    \begin{center}
        \includegraphics{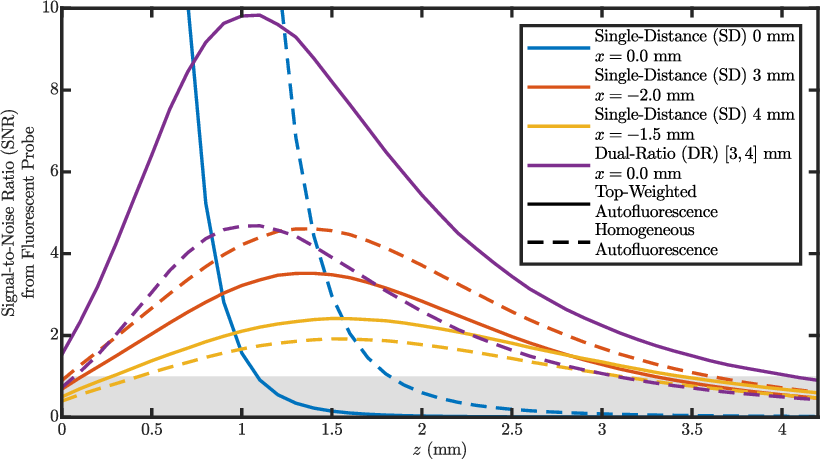}
    \end{center}
    \caption 
    { \label{fig:SNRvZ_SD}
    \Acrfull{SNR} from a fluorescent target below the centroid of the optodes used for each measurement type as a function of depth ($z$). Colors show different measurement types, and line type shows the \acrfull{AF} contributor distribution.\\
    Parameters: The same as Figure~\ref{fig:SNRmap} with assumed \SI{5}{\percent} Non-Cancelable noise. 
    } 
\end{figure}

The primary reason for this work is to explore if the \gls{DR} technique may be applied to \gls{DiFC}.
With this in mind, we must consider what metric to use to compare measurement methods such as \gls{SD} versus \gls{DR}.
Since the aim of \gls{DiFC} is to eventually non-invasively measure blood vessels of humans \textit{in-vivo}, the maximum depth that a method can detect a fluorescent target is the metric which should be used for this comparison.
One can determine this depth by examining Figure~\ref{fig:SNRmap} and finding the deepest colored region for each data type.
However, a line scan in $z$ of Figure~\ref{fig:SNRmap}, as shown in Figure~\ref{fig:SNRvZ_SD}, makes the relationships more apparent.
The $x$ values chosen in Figure~\ref{fig:SNRvZ_SD} were the centroid of the optodes (\textit{e.g.} $-2.0\hat{x}~\si{\milli\meter}$ for \texttt{1A}); thus, the maximum measurable depth can be found by looking at where the curves in Figure~\ref{fig:SNRvZ_SD} drop below a \gls{SNR} of one (the gray region).
Table~\ref{tab:maxDepth} summarizes these maximum measurable depths.
Note that a suitable vessel to measure in humans, such as the radial artery, is located approximately \SIrange{2}{4}{\milli\meter} deep\cite{selvaraj_ultrasoundevaluationeffect_2016}. 
\par

\begin{table}[htb]
    \centering
    \caption{Maximum Measurable Depths of Fluorescent Target for Different Measurement Types}\label{tab:maxDepth}
    \begin{tabular}{l|S[table-format=3.1]S[table-format=3.1]S[table-format=3.1]S[table-format=3.1]}
                                & \multicolumn{4}{c}{$\max[z[\text{\as{SNR}}>1]]$ (\si{\milli\meter})} \\
                                & {\as{SD} \as{rho}=\SI{0}{\milli\meter}}   & {\as{SD} \as{rho}=\SI{3}{\milli\meter}}   & {\as{SD} \as{rho}=\SI{4}{\milli\meter}}   & {\as{DR} \as{rho}s=[\num{3}, \num{4}]~\si{\milli\meter}} \\
                                \hline
                                \hline
        Surface-Weighted\\\as{AF} Contributors    & 1.0                                       & 3.2                                       & 3.4                                       & 4.0 \\
        \hline
        Homogeneous\\\as{AF} Contributors     & 1.8                                       & 3.6                                       & 3.1                                       & 3.0 \\
    \end{tabular}\\
    \tiny Acronyms: \Acrfull{SNR}, \acrfull{SD}, \acrfull{DR}, \acrfull{rho}, \& \acrfull{AF}. \\
    Parameters: The same as Figure~\ref{fig:SNRmap} with assumed \SI{5}{\percent} Non-Cancelable noise. 
\end{table}

As is evident in Table~\ref{tab:maxDepth}, the depth of \gls{DR} is heavily affected by the \gls{AF} contributor distribution. 
Specifically, if the \gls{AF} contributor distribution is surface-weighted, \gls{DR} is the best measurement type, while this is not true for homogeneous.
Evidence suggests that the \gls{AF} contributor distribution is surface-weighted \textit{in-vivo}. 
To investigate the spatial distribution of the \gls{AF} contributors, we measured the background signal of a mouse as described in Section~\ref{sec:meth:exp:miceaf}. 
The mean background with skin was \SI{17400\pm3500}{\nano\ampere}, whereas the mean background of the bare muscle (without skin) was \SI{9200\pm540}{\nano\ampere}, which is a reduction of \SI{47}{\percent}.
These results imply that \gls{DiFC} \gls{AF} is weighted to superficial tissue layers, and as discussed below, from what we can find in the literature, this seems to be the case qualitatively.
\par

A literature search did not reveal a specific quantification of the spatial distribution of \gls{AF} contributors.
Nevertheless, the available literature on known endogenous fluorophores strongly suggests that these should be weighted more heavily in the skin.
Specifically, \gls{NADH}, \gls{FAD}, collagen, and elastin fluoresce in the visible (violet and blue) wavelength regions\cite{monici_celltissueautofluorescence_2005, croce_autofluorescencespectroscopyimaging_2014} and to a lesser degree in the red and \gls{NIR} wavelength regions,\cite{hong_nearinfraredfluorophoresbiomedical_2017, gibbs_infraredfluorescenceimageguided_2012} the latter being most relevant for the potential use of \gls{DR} \gls{DiFC} in humans\cite{ivich_signalmeasurementconsiderations_2022}. 
Other significant contributors of tissue \gls{AF} in the red and \gls{NIR} wavelength regions are less obvious due to their rarity\cite{thomas_identifyingnovelendogenous_2016}.
Furthermore, \gls{AF} in the red and \gls{NIR} wavelength region may be more complex than visible \gls{AF} because of the overall reduction in optical scattering and absorption at longer wavelengths\cite{semenov_oxidationinducedautofluorescencehypothesis_2020}.
However, we found lipofuscin,\cite{marmorstein_spectralprofilingautofluorescence_2002} melanin,\cite{huang_cutaneousmelaninexhibiting_2006, wang_vivonearinfraredautofluorescence_2013} and heme-metabolic compounds like porphyrins and bilirubin,\cite{htun_nearinfraredautofluorescenceinduced_2017,lifante_roletissuefluorescence_2020} to be reported as auto-fluorescent in the red and \gls{NIR} wavelength region, all of which are present in the skin.
Bilirubin and other heme-metabolic products are mostly known for contributing to liver \gls{AF}\cite{demos_nearinfraredautofluorescenceimaging_2004}.
However, in the context of \gls{DiFC}, these compounds may accumulate in the skin from hematomas in the case of injury.
Thus, the available literature and experimental \gls{DiFC} measurements in mice (above) suggest that contributors to the background \gls{AF} in the red and \gls{NIR} wavelength region for potential \gls{DR} \gls{DiFC} are mostly in the skin as opposed to underlying tissue structures.
\par

\subsection{Consideration of Noise}\label{sec:dis:noise}
The noise and signal level used in these simulations was based on experimental data collected at \SI{3}{\milli\meter} \gls{SD} (Appendix~\ref{sec:app:expPar}). 
However, one noise parameter that cannot be derived from the collected experimental data is the correlation of the noise measured by different optodes.
This correlation between optode noises does not affect the \gls{SD} results but does significantly affect the \gls{DR} results since it affects how the \gls{DR} can cancel noise.
\par

\begin{figure}[ht]
    \begin{center}
        \includegraphics{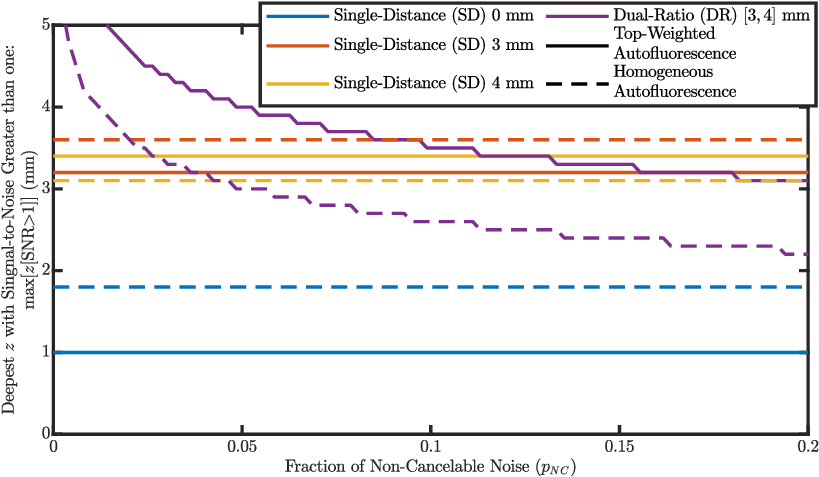}
    \end{center}
    \caption 
    { \label{fig:ZvsNonCanNoi}
    The deepest fluorescent target that each data type can measure (\acrfull{SNR} greater than one) as a function of the fraction of non-cancelable (by \acrfull{DR}) noise ($p_{NC}$). Colors show the data type, and line type shows the distribution of \acrfull{AF} contributors.\\
    Note: For a definition of see Appendix~\ref{sec:app:noise} or Equation~\ref{equ:CoupRand}.
    Parameters: The same as Figure~\ref{fig:SNRmap}. 
    } 
\end{figure}

We modeled noise as Gaussian random variables (\gls{Cr}) that multiplied the theoretical \gls{R} to yield the measured \gls{Ir} (Appendix~\ref{sec:app:noise}).
A separate independent random variable represents each optode (source or detector) and can be considered noise in the coupling, gain, power, or any other optode-specific noise.
This concept, explained in detail in Appendix~\ref{sec:app:expPar}, can be summarized by:
\begin{equation}\label{equ:CoupRand}
    \begin{split}
        \as{Ir}(\as{R},\as{sigmaI}^2)\simeq\hspace{10cm}\\
        \as{Cr}_{\text{Source}}\left(1,p_{\text{Source}}\left(\frac{\as{sigmaI}}{\as{R}}\right)^2\right)\times
        \as{Cr}_{\text{Detector}}\left(1,p_{\text{Detector}}\left(\frac{\as{sigmaI}}{\as{R}}\right)^2\right)\times\\
        \as{Cr}_{NC}\left(1,p_{NC}\left(\frac{\as{sigmaI}}{\as{R}}\right)^2\right)\times\as{R}
    \end{split}
\end{equation}
\noindent where, 
\begin{equation}
    p_{\text{Source}}+p_{\text{Detector}}+p_{NC}=1
\end{equation}
\noindent using the random variable notation where the first argument is the mean and the second the variance.
Since $\as{sigmaI}/\as{R}\ll1$, the $\as{Ir}$ defined by the right hand side of Equation~\ref{equ:CoupRand} is an excellent approximation of a random variable with mean $\as{R}$ and variance $\as{sigmaI}^2$ ($\as{Ir}(\as{R},\as{sigmaI}^2)$).
The \gls{DR} cancels all \glspl{Cr} associated with specific optodes (Equation~\ref{equ:app:DRwC}) but not the Non-Cancelable ($NC$) \gls{Cr}.
$NC$ is named such since it is the noise not canceled by \gls{DR}.
From this, we see that the parameter controlling the amount of noise that propagates into \gls{DR} is $p_{NC}$, which is the fraction of the variance which is non-cancelable by \gls{DR}.
\par

In all the above results in this work, $p_{NC}$ was assumed to be \SI{5}{\percent}.
Note that we do not know the physical origin of this noise if all \glspl{SD} are acquired simultaneously, so we apply this noise to noise introduced from multiplexing.
We expect that advanced multiplexing schemes may alleviate this noise (Section~\ref{sec:dis:implement}), but do not know what it is for the current system, so we assumed what we consider a reasonable value of \SI{5}{\percent}.
\par

To experiment with the effect of $p_{NC}$, we varied its value and determined the maximum measurable depth (where \gls{SNR} is greater than one) for each data type in Figure~\ref{fig:ZvsNonCanNoi}.
Note that all \gls{SD} measurements experience the same noise regardless of $p_{NC}$'s value, and only \gls{DR} is affected.
This is because the model was designed so that the total noise is the same regardless of the $p_{NC}$ value.
From Figure~\ref{fig:ZvsNonCanNoi}, we can find the allowable $p_{NC}$, which makes \gls{DR} sense deeper than the deepest \gls{SD} modeled.
For the simulation with surface-weighted \gls{AF} contributions, this is \SI{12}{\percent}, while for the homogeneous case, it is \SI{2}{\percent}.
This further emphasizes the advantage \gls{DR} has in the surface-weighted \gls{AF} medium. 
Additionally, this guides the amount of $p_{NC}$, which a system can have for \gls{DR} to have an advantage over \gls{SD}.
\par

\subsection{Future Implementation}\label{sec:dis:implement}
The experimental and computational data presented here were based on measurements using \gls{SD} or \gls{SR} (not presented for brevity) measurements \rl{taken in series} using our existing \gls{DiFC} instrument\cite{tan_vivoflowcytometry_2019, pace_nearinfrareddiffusevivo_2022}. 
In practice, implementation of a \gls{DR} for \gls{DiFC} would require the construction of a new \gls{DiFC} instrument capable of making simultaneous measurements with two sources and two detectors as in Figure~\ref{fig:introScheme}\textbf{(c)}.
Several optical designs would enable this.
For example, we could frequency encode the two laser sources by modulating them at different frequencies and de-modulating the fluorescence signals measured by the two detectors to separate the contributions from the two sources (frequency-multiplexing). 
Likewise, we could alternately illuminate the two laser sources (\texttt{S1} and \texttt{S2}) in an on-off pattern (time-multiplexing).
Assuming a peak width of \SI{10}{\milli\second} \textit{in-vivo}\cite{tan_vivoflowcytometry_2019}, this should be achievable by time-multiplexing the laser output at about \SI{1}{\kilo\hertz}.
\par

Fast time-multiplexing or frequency-multiplexing would be desirable because the \gls{DR} strategy would inherently cancel out signal drift and noise.
\Gls{DR} cancelable artifacts are associated with a multiplicative factor applied to a source or detector that does not change within the multiplexing cycle (Appendix~\ref{sec:app:noise:coup}).
This could include instrument drift or coupling of the sources and detectors to the skin surface.
However, this approach cannot cancel some sources of random noise in the signal (Appendix~\ref{sec:app:noise:coup} \& Figure~\ref{fig:ZvsNonCanNoi}).
In principle, using modulated laser sources and lock-in detection for frequency-multiplexing would improve the \gls{SNR} of detected peaks in the presence of non-cancellable noise.
Furthermore, as shown in Figures~\ref{fig:SIGsim}\&\ref{fig:expVsim}, the \gls{DR} \gls{DiFC} measurement using a parallel configuration (Figure~\ref{fig:introScheme}\textbf{(b)}) yielded a unique double-peak shape.
Therefore, we expect that using matched filters (or machine learning methods based on signal temporal shape) could improve peak detection.
\Gls{DR} noise would also be suppressed when the two \glspl{rho} are as different as possible\cite{blaney_designsourcedetector_2020}.
Therefore, custom-designed optical fiber bundles may be convenient for delivering and collecting light in a \gls{DR} \gls{DiFC} instrument.
For instance, we can design a new \gls{DiFC} optical fiber bundle that incorporates multiple source fibers with symmetric separations and large detector areas. 
\par

Finally, we note that this work is the first case of \rl{dual-slope} / \gls{DR} applied to fluorescence in general, not only to \gls{DiFC}.
\rl{dual-slope} / \gls{DR} was first developed for \gls{NIRS} and measurement of \gls{dmua}\cite{sassaroli_dualslopemethodenhanced_2019, blaney_phasedualslopesfrequencydomain_2020}.
Now, with the methods presented in this work, we believe \rl{dual-slope} / \gls{DR} may also be applied to diffuse fluorescence spectroscopy and tomography\cite{htun_nearinfraredautofluorescenceinduced_2017,grosenick_reviewopticalbreast_2016, croce_autofluorescencespectroscopyimaging_2014,wang_deeptissuefocalfluorescence_2012,klose_inversesourceproblem_2005,georgakoudi_trimodalspectroscopydetection_2002}.
\rl{dual-slope} / \gls{DR} should lend itself well to any application which aims to measure changes in fluorescence (\textit{i.e.}, instead of \gls{dmua} as it was used for in \gls{NIRS}).
Thus, we open the door to future work regarding \rl{dual-slope} / \gls{DR} fluorescence.

\section{Conclusion}\label{sec:con}
This work explored the feasibility of using \gls{DR} for \gls{DiFC}. 
We accomplished this by running \gls{MC} models of the expected \gls{DR} \gls{SNR} based on noise and fluorescence parameters extracted from experimental \gls{DiFC} data.
From this exploration, we modify two key factors which control whether \gls{DR} is beneficial to \gls{DiFC} or not.
The first is the distribution of \gls{AF} contributors \textit{in-vivo}, with surface-weighted giving \gls{DR} methods an advantage. 
Experiments on mice and a literature search suggest that the \textit{in-vivo} \gls{AF} contributor distribution is concentrated in the skin, suggesting that \gls{DR} will have an advantage over traditional \gls{SD} methods.
The second key parameter is the portion of non-cancellable (by \gls{DR}) noise in the measurement.
This noise is not associated with multiplicative optode factors (\textit{e.g.} tissue coupling, source power, or detector efficiency), which are constant within a multiplexing cycle.
A \gls{DiFC} instrument designed for \gls{DR} would need to be built to explore and address this. 
\rl{A future \gls{DiFC} instrument may achieve the \gls{DR} measurement by time- or frequency-multiplexing leading to the required level of non-cancelable noise.
Therefore, this work laid the groundwork to identify the key parameters of concern and aid in designing a \gls{DR} \gls{DiFC} instrument.}
\par

\appendix 
\glsresetall

\section{Definition of Measurement Types}\label{sec:app:defs}
\subsection{Single-Distance}\label{sec:app:defs:SD}
The raw measurement from a \gls{DiFC} system is the \gls{I} measured between a source and a detector which we also refer to as the \gls{SD} here. 
Using this raw measurement we define the \gls{dSD} as the background-subtracted \gls{I} as follows:
\begin{equation}\label{equ:app:dSD}
    \acrshort{dSD}(\acrshort{t})=\acrshort{I}(t)-\acrshort{I}_0
\end{equation}
\noindent where, \gls{dSD} is expressed as a function of \gls{t} and $\acrshort{I}_0$ is the average or \gls{I} over a baseline \gls{t}.
\rl{$\acrshort{I}_0$ could be found in various ways, for example: based on an initial measurement or an average over a preceding time period, regardless \gls{dSD} represents a change in \gls{I} over \gls{t}.}
Here, the primary measurement parameter is the \gls{rho}, in this work \glspl{SD} are evaluated at \gls{rho} of either \SI{3}{\milli\meter} or \SI{4}{\milli\meter}.

\subsection{Single-Ratio}
Next we define the \gls{SR} which is a ratio of \glspl{I} measured at long and short \glspl{rho}:
\begin{equation}\label{equ:app:SR}
    \acrshort{SR}=\frac{\acrshort{I}_l}{\acrshort{I}_s}
\end{equation}
\noindent where, the $l$ or $s$ subscripts represent measurements at either long or short \glspl{rho}, respectively.
Using Equation~\ref{equ:app:SR} we next express the \gls{dSR}:
\begin{equation}\label{equ:app:dSR}
    \acrshort{dSR}(\acrshort{t})=\frac{\acrshort{I}_l(\acrshort{t})}{\acrshort{I}_s(\acrshort{t})}-\frac{\acrshort{I}_{l,0}}{\acrshort{I}_{s,0}}
\end{equation}
\noindent which, as with \gls{dSD}, is also expressed as a function of time.
The primary parameters for \gls{SR} are the long and short \glspl{rho} which are \SI{3}{\milli\meter} and \SI{4}{\milli\meter} in this work.

\subsection{Dual-Ratio}\label{sec:app:defs:DR}
Finally, we define the \gls{DR}, which is the geometric mean of two \glspl{SR} related through geometric requirements. 
These requirements can be summarized by stating that optodes that contribute to the short \gls{rho} for one \gls{SR} contribute to the long \gls{rho} for the other \gls{SR} in a \gls{DR}; and vice-versa.
Suppose we have two \glspl{SR} named $\acrshort{SR}_{\text{\texttt{I}}}$ and $\acrshort{SR}_{\text{\texttt{II}}}$.
Naming sources as numbers and detectors as letters, we can say that $\acrshort{SR}_{\text{\texttt{I}}}$ contains one detector (\texttt{A}) and two sources (\texttt{1} \& \texttt{2}) such that the short \gls{rho} is obtained between \texttt{1} \& \texttt{A} and the long \gls{rho} between \texttt{2} \& \texttt{A} (Figure~\ref{fig:introScheme}\textbf{(d)}).
Now consider $\acrshort{SR}_{\text{\texttt{II}}}$ to be made of detector \texttt{B} and the same two sources \texttt{1} \& \texttt{2} with the short \gls{rho} being between \texttt{B} \& \texttt{2} and the long \gls{rho} between \texttt{B} \& \texttt{1}. 
Now notice that for $\acrshort{SR}_{\text{\texttt{I}}}$ \texttt{1} forms the short \gls{rho} but for $\acrshort{SR}_{\text{\texttt{II}}}$ \texttt{1} forms the long \gls{rho}.
Similarly for $\acrshort{SR}_{\text{\texttt{II}}}$ \texttt{2} forms the short \gls{rho} but for $\acrshort{SR}_{\text{\texttt{I}}}$ \texttt{2} forms the long \gls{rho}. 
Therefore the geometric requirements are satisfied, and the geometric mean of $\acrshort{SR}_{\text{\texttt{I}}}$ \& $\acrshort{SR}_{\text{\texttt{II}}}$ form a \gls{DR}:
\begin{equation}\label{equ:app:DR}
    \acrshort{DR}=\sqrt{\acrshort{SR}_{\text{\texttt{I}}}\acrshort{SR}_{\text{\texttt{II}}}}=\sqrt{\frac{\acrshort{I}_{l,\text{\texttt{I}}}\acrshort{I}_{l,\text{\texttt{II}}}}{\acrshort{I}_{s,\text{\texttt{I}}}\acrshort{I}_{s,\text{\texttt{II}}}}}
\end{equation}
\noindent and thus \gls{dDR} can we written as:
\begin{equation}\label{equ:app:dDR}
    \acrshort{dDR}(\acrshort{t})=
    \sqrt{\frac{\acrshort{I}_{l,\text{\texttt{I}}}(\acrshort{t})\acrshort{I}_{l,\text{\texttt{II}}}(\acrshort{t})}{\acrshort{I}_{s,\text{\texttt{I}}}(\acrshort{t})\acrshort{I}_{s,\text{\texttt{II}}}(\acrshort{t})}}-
    \sqrt{\frac{\acrshort{I}_{l,\text{\texttt{I}},0}\acrshort{I}_{l,\text{\texttt{II}},0}}{\acrshort{I}_{s,\text{\texttt{I}},0}\acrshort{I}_{s,\text{\texttt{II}},0}}}
\end{equation}
\noindent Similarly to \gls{SR} the parameters that are important for \gls{DR} are the short and long \glspl{rho}. However, in this case, there are two short and two long \glspl{rho}. In this work both short are considered to be \SI{3}{\milli\meter} and both long \SI{4}{\milli\meter}.
\par

\section{Theory}\label{sec:app:theo}
\subsection{Modeling Fluorescent Reflectance}
To model the detected fluorescent \gls{R} we must consider two processes. 
First, the delivery of power to the fluorophores. 
Second, the transport of emitted light from the fluorophores to the detector. 
\par

\subsubsection{Dependence on Fluorophore Position}
The expression for the \gls{Q} absorbed by a fluorophore ($\acrshort{Q}_{fl,ab}(\acrshort{r}_{fl})$; unit of \si{\milli\watt\per\milli\meter\cubed}) at the \gls{r} of the fluorophore ($\acrshort{r}_{fl}$) can be written as:
\begin{equation}
    \acrshort{Q}_{fl,ab}(\acrshort{r}_{fl})=\acrshort{PHI}(\acrshort{r}_{src}\shortrightarrow\acrshort{r}_{fl}){\acrshort{mua}}_{,fl}(\acrshort{r}_{fl})
\end{equation}
\noindent where, ${\acrshort{mua}}_{,fl}$ is the \gls{mua} for the fluorophore (unit of \si{\per\milli\meter}) and $\acrshort{PHI}(\acrshort{r}_{src}\shortrightarrow\acrshort{r}_{fl})$ is the \gls{PHI} (using the excitation \gls{lam} optical properties) from a pencil beam at the source position ($\acrshort{r}_{src}$) to the fluorophore (unit of \si{\milli\watt\per\milli\meter\squared}).
Next, we consider the \gls{Q} emitted by the fluorophore ($\acrshort{Q}_{fl,em}(\acrshort{r}_{fl})$; unit of \si{\milli\watt\per\milli\meter\cubed}):
\begin{equation}\label{equ:app:Qflem}
    \acrshort{Q}_{fl,em}(\acrshort{r}_{fl})=\acrshort{Q}_{fl,ab}(\acrshort{r}_{fl})\acrshort{eta}(\acrshort{r}_{fl})
\end{equation}
\noindent where, \gls{eta} represents the proportion of absorbed power converted to fluorescence.
Thinking of the medium as voxelized with voxels with \gls{V}, we volumetrically integrate Equation~\ref{equ:app:Qflem} over \gls{V} (unit of \si{\milli\meter\cubed}) to yield the \gls{P} emitted by the fluorophore at position $\acrshort{r}_{fl}$ inside the voxel ($\acrshort{P}_{fl,em}(\acrshort{r}_{fl})$; unit of \si{\milli\watt}):
\begin{equation}
    \acrshort{P}_{fl,em}(\acrshort{r}_{fl})=\acrshort{Q}_{fl,ab}(\acrshort{r}_{fl})\acrshort{eta}(\acrshort{r}_{fl})V
\end{equation}
\noindent Finally, we find the fluorescent \gls{R} from the fluorophore within $V$ at $\acrshort{r}_{fl}$ which is excited by a pencil beam at $\acrshort{r}_{src}$ and detected at $\acrshort{r}_{det}$ ($\acrshort{R}_{fl}(\acrshort{r}_{src}\shortrightarrow\acrshort{r}_{fl}\shortrightarrow\acrshort{r}_{det})$; unit of \si{\milli\watt\per\milli\meter\squared}). $\acrshort{R}_{fl}(\acrshort{r}_{src}\shortrightarrow\acrshort{r}_{fl}\shortrightarrow\acrshort{r}_{det})$ is obtained by multiplying $\acrshort{P}_{fl,em}(\acrshort{r}_{fl})$ by the \gls{R} Green's function  (using the emission \gls{lam} optical properties) from $\acrshort{r}_{fl}$ to $\acrshort{r}_{det}$ ($\acrshort{R}_{\text{Green}}(\acrshort{r}_{fl}\shortrightarrow\acrshort{r}_{det})$; unit of \si{\per\milli\meter\squared}):
\begin{equation}
    \acrshort{R}_{fl}(\acrshort{r}_{src}\shortrightarrow\acrshort{r}_{fl}\shortrightarrow\acrshort{r}_{det})=\acrshort{P}_{fl,em}(\acrshort{r}_{fl})\acrshort{R}_{\text{Green}}(\acrshort{r}_{fl}\shortrightarrow\acrshort{r}_{det})
\end{equation}
\par

To simplify these expressions, we rewrite them to yield:
\begin{equation}
    \acrshort{R}_{fl}(\acrshort{r}_{src}\shortrightarrow\acrshort{r}_{fl}\shortrightarrow\acrshort{r}_{det})=\acrshort{PHI}(\acrshort{r}_{src}\shortrightarrow\acrshort{r}_{fl}) \acrshort{R}_{\text{Green}}(\acrshort{r}_{fl}\shortrightarrow\acrshort{r}_{det})V {\acrshort{mua}}_{,fl}(\acrshort{r}_{fl})\acrshort{eta}(\acrshort{r}_{fl})
\end{equation}
\noindent Note that $\acrshort{R}_{\text{Green}}(\acrshort{r}_{fl}\shortrightarrow\acrshort{r}_{det})$ is rather difficult to calculate using methods such as Monte-Carlo, so we may use the Green's function of the \gls{PHI} (using the emission \gls{lam} optical properties) with a source placed at $\acrshort{r}_{det}$ ($\acrshort{PHI}_{\text{Green}}(\acrshort{r}_{det}\shortrightarrow\acrshort{r}_{fl})$; unit of \si{\per\milli\meter\squared}) as an approximation according to the adjoint method\cite{Yao_Biomed.Opt.ExpressBOE18_DirectApproach}. 
Additionally, we note that $\acrshort{PHI}(\acrshort{r}_{src}\shortrightarrow\acrshort{r}_{fl})$ can be expressed as the Green's function for \gls{PHI} ($\acrshort{PHI}_{\text{Green}}(\acrshort{r}_{src}\shortrightarrow\acrshort{r}_{fl})$; unit of \si{\per\milli\meter\squared}) multiplied by the \gls{P} of the source ($\acrshort{P}_{src}$; unit of \si{\milli\watt}). These approximations and substitutions yield:
\begin{equation}
    \acrshort{R}_{fl}(\acrshort{r}_{src}\shortrightarrow\acrshort{r}_{fl}\shortrightarrow\acrshort{r}_{det})\approx\acrshort{P}_{src} \acrshort{PHI}_{\text{Green}}(\acrshort{r}_{src}\shortrightarrow\acrshort{r}_{fl}) \acrshort{PHI}_{\text{Green}}(\acrshort{r}_{det}\shortrightarrow\acrshort{r}_{fl}) V {\acrshort{mua}}_{,fl}(\acrshort{r}_{fl})\acrshort{eta}(\acrshort{r}_{fl})
\end{equation}
\noindent We remind that $\acrshort{PHI}_{\text{Green}}(\acrshort{r}_{src}\shortrightarrow\acrshort{r}_{fl})$ and $\acrshort{PHI}_{\text{Green}}(\acrshort{r}_{det}\shortrightarrow\acrshort{r}_{fl})$ are the Green's functions calculated at the excitation and emission wavelengths, respectively.
In this work, however, we assume the optical properties at the two wavelengths to be the same.
Last we define the \gls{W}\cite{ivich_signalmeasurementconsiderations_2022} which is a helpful parameter in understanding how the $\acrshort{R}_{fl}(\acrshort{r}_{src}\shortrightarrow\acrshort{r}_{fl}\shortrightarrow\acrshort{r}_{det})$ depends on the spatial location of the fluorophore.
\begin{equation}\label{equ:app:W}
    \acrshort{W}(\acrshort{r}_{src}\shortrightarrow\acrshort{r}_{fl}\shortrightarrow\acrshort{r}_{det})=\acrshort{PHI}_{\text{Green}}(\acrshort{r}_{src}\shortrightarrow\acrshort{r}_{fl}) \acrshort{PHI}_{\text{Green}}(\acrshort{r}_{det}\shortrightarrow\acrshort{r}_{fl}) V
\end{equation}
\begin{equation}\label{equ:app:Rfl}
    \acrshort{R}_{fl}(\acrshort{r}_{src}\shortrightarrow\acrshort{r}_{fl}\shortrightarrow\acrshort{r}_{det})\approx \acrshort{P}_{src} \acrshort{W}(\acrshort{r}_{src}\shortrightarrow\acrshort{r}_{fl}\shortrightarrow\acrshort{r}_{det}) {\acrshort{mua}}_{,fl}(\acrshort{r}_{fl})\acrshort{eta}(\acrshort{r}_{fl})
\end{equation}
\noindent In-fact if ${\acrshort{mua}}_{,fl}(\acrshort{r}_{fl})$ and $\acrshort{eta}(\acrshort{r}_{fl})$ are both spatially independent, $\acrshort{W}(\acrshort{r}_{src}\shortrightarrow\acrshort{r}_{fl}\shortrightarrow\acrshort{r}_{det})$ (unit of \si{\per\milli\meter}) is directly proportional to $\acrshort{R}_{fl}(\acrshort{r}_{src}\shortrightarrow\acrshort{r}_{fl}\shortrightarrow\acrshort{r}_{det})$ and representative of how the spatial location of the fluorophore contributes to the detected fluorescence.
\par

\subsubsection{Fluorescent Reflectance from Multiple Fluorophores}
It is assumed that the \gls{R} measured from each voxel containing a fluorophore may be linearly combined. Therefore:
\begin{equation}\label{equ:app:hetBck}
    \acrshort{R}_{fl}(\acrshort{r}_{src}\shortrightarrow\Sigma\acrshort{r}_{fl}\shortrightarrow\acrshort{r}_{det})\approx \acrshort{P}_{src} \Sigma_{i} \acrshort{W}(\acrshort{r}_{src}\shortrightarrow\acrshort{r}_{fl,i}\shortrightarrow\acrshort{r}_{det}) {\acrshort{mua}}_{,fl}(\acrshort{r}_{fl,i})\acrshort{eta}(\acrshort{r}_{fl,i})
\end{equation}
\noindent and if ${\acrshort{mua}}_{,fl}(\acrshort{r}_{fl,i})$ and $\acrshort{eta}(\acrshort{r}_{fl,i})$ \rl{(where the subscript $i$ indicates the $i$\textsuperscript{th} voxel)} are spatially constant:
\begin{equation}\label{equ:app:homBck}
    \acrshort{R}_{fl}(\acrshort{r}_{src}\shortrightarrow\Sigma\acrshort{r}_{fl}\shortrightarrow\acrshort{r}_{det})\approx \acrshort{P}_{src} {\acrshort{mua}}_{,fl}\acrshort{eta} \acrshort{W}(\acrshort{r}_{src}\shortrightarrow\Sigma\acrshort{r}_{fl}\shortrightarrow\acrshort{r}_{det})
\end{equation}
\noindent where,
\begin{equation}
    \acrshort{W}(\acrshort{r}_{src}\shortrightarrow\Sigma\acrshort{r}_{fl}\shortrightarrow\acrshort{r}_{det})= V \Sigma_i \acrshort{PHI}_{\text{Green}}(\acrshort{r}_{src}\shortrightarrow\acrshort{r}_{fl,i}) \acrshort{PHI}_{\text{Green}}(\acrshort{r}_{det}\shortrightarrow\acrshort{r}_{fl,i})
\end{equation}
\par

The background \gls{R} is a case where the sum of all the \gls{AF} fluorophores from every voxel must be considered. 
In this case, Equation~\ref{equ:app:hetBck} is considered for a heterogeneous distribution of \gls{AF} \gls{mua} or \gls{eta} (\textit{i.e.}, \gls{AF} contributions) and Equation~\ref{equ:app:homBck} for a homogeneous \gls{AF} \gls{mua} or \gls{eta}. 
Any signal from a target fluorescent target is then added to this background \gls{R} using Equation~\ref{equ:app:Rfl}.
This yields an expression for the fluorescent \gls{R} considering contributions from both the target fluorescent target at $\acrshort{r}_{target}$ and all \gls{AF} contributors:
\begin{equation}\label{equ:app:Rtarget}
    \acrshort{R}(\acrshort{r}_{target})\approx
         \acrshort{P}_{src} \left(
         \Sigma_{i} \acrshort{W}(\acrshort{r}_{i}) {\acrshort{mua}}_{,\text{\acrshort{AF}}}(\acrshort{r}_{i})\acrshort{eta}_{\text{\acrshort{AF}}}(\acrshort{r}_{i})+
         \acrshort{W}(\acrshort{r}_{target}) {\acrshort{mua}}_{,target}(\acrshort{r}_{target})\acrshort{eta}_{target}(\acrshort{r}_{target})
         \right)
\end{equation}
\noindent Here we have dropped the notation showing the source and detector \gls{r} for brevity. Notice that the background \gls{R} expressed as $\acrshort{R}_0$ is the \gls{AF} term, such that:
\begin{equation}\label{equ:app:R0}
    \acrshort{R}_0\approx
         \acrshort{P}_{src}
         \Sigma_{i} \acrshort{W}(\acrshort{r}_{i}) {\acrshort{mua}}_{,\text{\acrshort{AF}}}(\acrshort{r}_{i})\acrshort{eta}_{\text{\acrshort{AF}}}(\acrshort{r}_{i})
\end{equation}
and,
\begin{equation}\label{equ:app:Rsim}
    \acrshort{R}(\acrshort{r}_{target})\approx
         \acrshort{P}_{src} \acrshort{W}(\acrshort{r}_{target}) {\acrshort{mua}}_{,target}(\acrshort{r}_{target})\acrshort{eta}_{target}(\acrshort{r}_{target}) + \acrshort{R}_0
\end{equation}
\noindent For simulations these expressions, namely Equation~\ref{equ:app:Rtarget}~\&~\ref{equ:app:R0}, are used as the simulated \glspl{R} to obtained simulated measured \gls{I} using methods in Appendix~\ref{sec:app:noise}. These simulated measured \glspl{I} can then be used with the expressions in Appendix~\ref{sec:app:defs} to obtain simulated measurement types \gls{SD} and \gls{DR}.

\subsection{Simulating Noise \& Coupling Coefficients}\label{sec:app:noise}
\subsubsection{Coupling Coefficients}\label{sec:app:noise:coup}
We assume that the measured \gls{I} is related to the theoretical \gls{R} through multiplicative \glspl{C} typically associated with each optode.
Notice that we distinguish between the measured and theoretical values by naming the former \gls{I} and the latter \gls{R}.
As such, the equations in Appendix~\ref{sec:app:defs} show measurement types derived from measured (or simulated measured) \glspl{I} while the expressions in Appendix~\ref{sec:app:theo} show calculation of theoretical \glspl{R}.
The use of \glspl{C} in this section serves to connect \gls{I} and \gls{R}.
Therefore, \gls{C} may take the proper units to convert the measured \gls{I} (measured in \si{\nano\ampere} from a \gls{PMT} for example) to the unit of \gls{R} (\si{\milli\watt\per\milli\meter\squared}).
Additionally, if noise is considered, \gls{C} may be a random variable representing the noise that confounds \gls{I}.
\par

Given the optode arrangement in Figure~\ref{fig:introScheme}\textbf{(c)} we have two sources (\texttt{1} \& \texttt{2}) and two detectors (\texttt{A} \& \texttt{B}). 
Therefore, four measurements of \gls{I} between a source and detector are possible: $\as{I}_{\text{\texttt{1A}}}$, $\as{I}_{\text{\texttt{1B}}}$, $\as{I}_{\text{\texttt{2A}}}$, and $\as{I}_{\text{\texttt{2B}}}$. Which are related to \gls{R} through \glspl{C} as follows:
\begin{equation}\label{equ:app:R2I_1A}
    \as{I}_{\text{\texttt{1A}}}=\as{C}_{\text{\texttt{1}}}\as{C}_{\text{\texttt{A}}}\as{C}_{NC,\alpha}\as{R}_{\text{\texttt{1A}}}
\end{equation}
\begin{equation}
    \as{I}_{\text{\texttt{1B}}}=\as{C}_{\text{\texttt{1}}}\as{C}_{\text{\texttt{B}}}\as{C}_{NC,\beta}\as{R}_{\text{\texttt{1B}}}
\end{equation}
\begin{equation}
    \as{I}_{\text{\texttt{2A}}}=\as{C}_{\text{\texttt{2}}}\as{C}_{\text{\texttt{A}}}\as{C}_{NC,\gamma}\as{R}_{\text{\texttt{2A}}}
\end{equation}
\begin{equation}
    \as{I}_{\text{\texttt{2B}}}=\as{C}_{\text{\texttt{2}}}\as{C}_{\text{\texttt{B}}}\as{C}_{NC,\delta}\as{R}_{\text{\texttt{2B}}}
\end{equation}
\noindent where, the \glspl{C} with \texttt{1}, \texttt{2}, \texttt{A}, or \texttt{B} subscripts are associated with the corresponding optode (source or detector). 
The $\as{C}_{NC}$s are Non-Cancelable coupling factors or noise not associated with a specific optode and from an unknown source. 
Non-Cancelable refers to these $\as{C}_{NC}$s not being canceled by the \gls{DR} measurement, as is shown later in this section.
Notice that there is also a unique $\as{C}_{NC}$ (shown with Greek subscripts) for each \gls{I} measurement.
\par

\subsubsection{Noise}
Now we consider noise and thus model the \glspl{C} as Gaussian \glspl{Cr}. 
In the following, we assume \gls{I} and \gls{R} have the same units and scale (making \gls{Cr} unit-less and mean one); however, in reality, a coefficient converting \gls{R} to \gls{I} \rl{multiplies} the \glspl{Cr}.
We assume that the \gls{sigmaI} is proportional to the \gls{R} such that \gls{sigmaRel} is constant across different \glspl{I}, where:
\begin{equation}\label{equ:app:sigmaRel}
    \as{sigmaRel}=\as{sigmaI}/\as{R}
\end{equation}
\noindent Therefore, the \gls{Ir} from Equation~\ref{equ:CoupRand} has the mean value \gls{R} and variance $\as{sigmaI}^2$ ($\as{Ir}(\as{R},\as{sigmaI}^2)=\as{Ir}(\as{R},\as{sigmaRel}^2\as{R}^2)$).
This can be modeled by taking the mean of all \glspl{Cr} one and the sum of the three \glspl{Cr}' variance who contribute to a single \gls{I} equal to $\as{sigmaI}^2/\as{R}^2$.
To model this, we introduce two parameters, the fraction of noise contributed from each optode ($p_{opt}$) and the fraction of Non-Cancelable noise ($p_{NC}$). 
Thus we can model the \gls{I} noise as:
\begin{equation}\label{equ:app:I1AcoupRand}
    \as{Ir}_{\text{\texttt{1A}}}(\as{R}_{\text{\texttt{1A}}},{\as{sigmaI}}_{\text{\texttt{1A}}}^2)=\as{Cr}_{\text{\texttt{1}}}(1,p_{opt}\as{sigmaRel}^2)\as{Cr}_{\text{\texttt{A}}}(1,p_{opt}\as{sigmaRel}^2)\as{Cr}_{NC,\alpha}(1,p_{NC}\as{sigmaRel}^2)\as{R}_{\text{\texttt{1A}}}
\end{equation}
\begin{equation}
    \as{Ir}_{\text{\texttt{1B}}}(\as{R}_{\text{\texttt{1B}}},{\as{sigmaI}}_{\text{\texttt{1B}}}^2)=\as{Cr}_{\text{\texttt{1}}}(1,p_{opt}\as{sigmaRel}^2)\as{Cr}_{\text{\texttt{B}}}(1,p_{opt}\as{sigmaRel}^2)\as{Cr}_{NC,\beta}(1,p_{NC}\as{sigmaRel}^2)\as{R}_{\text{\texttt{1B}}}
\end{equation}
\begin{equation}
    \as{Ir}_{\text{\texttt{2A}}}(\as{R}_{\text{\texttt{2A}}},{\as{sigmaI}}_{\text{\texttt{2A}}}^2)=\as{Cr}_{\text{\texttt{2}}}(1,p_{opt}\as{sigmaRel}^2)\as{Cr}_{\text{\texttt{A}}}(1,p_{opt}\as{sigmaRel}^2)\as{Cr}_{NC,\gamma}(1,p_{NC}\as{sigmaRel}^2)\as{R}_{\text{\texttt{2A}}}
\end{equation}
\begin{equation}
    \as{Ir}_{\text{\texttt{2B}}}(\as{R}_{\text{\texttt{2B}}},{\as{sigmaI}}_{\text{\texttt{2B}}}^2)=\as{Cr}_{\text{\texttt{2}}}(1,p_{opt}\as{sigmaRel}^2)\as{Cr}_{\text{\texttt{B}}}(1,p_{opt}\as{sigmaRel}^2)\as{Cr}_{NC,\delta}(1,p_{NC}\as{sigmaRel}^2)\as{R}_{\text{\texttt{2B}}}
\end{equation}
\noindent where,
\begin{equation}
    2p_{opt}+p_{NC}=1
\end{equation}
\noindent In the simulations shown in this work, the \glspl{Cr} takes the form of a Gaussian distribution.

\subsubsection{Propagation of Coupling Coefficients and Noise to Dual-Ratio}
This work presents the results of \gls{SD} and \gls{DR}. 
Therefore, it is helpful to understand how these noise models affect each.
For the case of \gls{SD}, coupling and noise affect the measurement in the same way as \gls{I} is shown to be involved above.
However, for \gls{DR} all coupling and noise are canceled except for the Non-Cancelable component.
To show this we rewrite Equation~\ref{equ:app:DR} considering the optodes in Figure~\ref{fig:introScheme}\textbf{(d)} but replacing \glspl{I} with \glspl{C} and \glspl{R}:

\begin{equation}\label{equ:app:DRwC}
    \acrshort{DR}=
        \sqrt{\frac{
        \as{C}_{\text{\texttt{1}}}\as{C}_{\text{\texttt{B}}}\as{C}_{NC,\beta}\as{R}_{\text{\texttt{1B}}}
        \as{C}_{\text{\texttt{2}}}\as{C}_{\text{\texttt{A}}}\as{C}_{NC,\gamma}\as{R}_{\text{\texttt{2A}}}
        }{
        \as{C}_{\text{\texttt{1}}}\as{C}_{\text{\texttt{A}}}\as{C}_{NC,\alpha}\as{R}_{\text{\texttt{1A}}}
        \as{C}_{\text{\texttt{2}}}\as{C}_{\text{\texttt{B}}}\as{C}_{NC,\delta}\as{R}_{\text{\texttt{2B}}}
        }}=
        \sqrt{\frac{
        \as{C}_{NC,\beta}
        \as{C}_{NC,\gamma}
        }{
        \as{C}_{NC,\alpha}
        \as{C}_{NC,\delta}
        }}
        \sqrt{\frac{
        \as{R}_{\text{\texttt{1B}}}
        \as{R}_{\text{\texttt{2A}}}
        }{
        \as{R}_{\text{\texttt{1A}}}
        \as{R}_{\text{\texttt{2B}}}
        }}
\end{equation}
leading to,
\begin{equation}
    \acrshort{DR}=
        \sqrt{\frac{
        \as{C}_{NC,\beta}
        \as{C}_{NC,\gamma}
        }{
        \as{C}_{NC,\alpha}
        \as{C}_{NC,\delta}
        }}
        \as{DR}_{theo}
\end{equation}
\noindent where, $\as{DR}_{theo}$ is the theoretical \gls{DR}. 
Therefore, the measured and theoretical \gls{DR} are equal if $p_{NC}=0$. 
Additionally, this demonstrates that \gls{DR} eliminates \glspl{C} associated with multiplicative factors belonging to the optodes, and only the Non-Cancelable $\as{C}_{NC}$s remain in the \gls{DR} measurement.

\section{Simulation Parameters from Experimental Data}\label{sec:app:expPar}
\subsection{Determining Background and Peak Fluorescence Coefficients}
The experimental data from Section~\ref{sec:meth:exp:phantom} was used to calculate \gls{eta} and \gls{mua} for both background and target ($\as{eta}_{\as{AF}}{\as{mua}}_{,\as{AF}}$ and $\as{eta}_{target}{\as{mua}}_{,target}$, respectively). 
This allowed for the simulated measurement which we present.
The \glspl{PMT} output measurement was \gls{I} in \si{\nano\ampere}; therefore, it was assumed that the \glspl{C} in Equation~\ref{equ:app:R2I_1A} together had the units of \si{\watt\per\milli\meter\squared\per\nano\ampere} making \gls{R} have the unit of \si{\watt\per\milli\meter\squared}.
We assumed this value to be \SI{271e-15}{\watt\per\milli\meter\squared\per\nano\ampere} based on detector gain (\num{1e4}), detector area (\SI{0.565}{\milli\meter\squared}), and \gls{lam} (\SI{810}{\nano\meter}).
However, since this value cancels whenever a unit-less quantity is calculated (such as \gls{SNR}), we stress that what we assume for this value does not matter for the results; we simply assume one to match units between the experimental data and \gls{MC}.
As for \gls{MC} parameters for calculation of \gls{W}, we assume a \gls{mua} of \SI{0.002}{\per\milli\meter}, a \gls{mus} of \rl{\SI{7}{\per\milli\meter}}, and a \gls{n} of \num{1.37}.
The experimental data collected for these calculations were measured at a \gls{rho} of \SI{3}{\milli\meter} and a source \gls{P} of \SI{75}{\milli\watt}.
Finally, the \gls{MC} used a voxel size of $\SI{0.1}{\milli\meter}\times\SI{0.1}{\milli\meter}\times\SI{0.1}{\milli\meter}$ and launched \num{1e9} photons.
\par

\subsubsection{Background Autofluorescence}\label{sec:app:expPar:bck}
To find the background \gls{I} / \gls{R}, the moving mean using a \SI{1}{\second} window (with a sample rate of \SI{2}{\kilo\hertz}) was found for all experimental data, with outliers removed (defining outliers as values more than three scaled mean absolute deviations from the median).
Then, the median of all moving mean values was found and taken as the background.
Using this method, we found an average measured background at $\as{rho}=\SI{3}{\milli\meter}$, of \SI{170}{\nano\ampere} or \SI{46.1}{\pico\watt\per\milli\meter\squared}.
\par

When assuming the \gls{AF} contributor distribution to be heterogeneous and surface-weighted, we model the $\as{eta}_{\as{AF}}$ as exponentially decaying as one goes deeper into the medium:
\begin{equation}
    \as{eta}_{\as{AF},het}(\as{r})=e^{\ln{(1/2)}(\as{r}\cdot\hat{z})/\SI{0.1}{\milli\meter}}
\end{equation}
\noindent Then rewrite Equation~\ref{equ:app:R0} as follows:
\begin{equation}
    \as{R}_{0} \approx
        \as{P}_{src} {\as{mua}}_{,\as{AF},het}
        \Sigma_{\text{All \as{r}}} \as{W}(\as{r}) e^{\ln{(1/2)}(\as{r}\cdot\hat{z})/\SI{0.1}{\milli\meter}}
\end{equation}
\noindent and solve for ${\as{mua}}_{,\as{AF}}$:
\begin{equation}
    {\as{mua}}_{,\as{AF},het} \approx \frac{\as{R}_{0}}
        {\as{P}_{src} \Sigma_{\text{All \as{r}}} \as{W}(\as{r}) e^{\ln{(1/2)}(\as{r}\cdot\hat{z})/\SI{0.1}{\milli\meter}}}
\end{equation}
\noindent With this method we found that ${\as{mua}}_{,\as{AF},het} \as{eta}_{\as{AF},het}=\SI{228e-9}{\per\milli\meter}\times e^{\ln{(1/2)}(\as{r}\cdot\hat{z})/\SI{0.1}{\milli\meter}}$ for the measured background \gls{AF}.
\par

In the case of homogeneous \gls{AF} contributors we instead rewrite and solve Equation~\ref{equ:app:R0} as:
\begin{equation}
    {\as{mua}}_{,\as{AF},hom} \as{eta}_{\as{AF},hom} \approx \frac{\as{R}_{0}}
        {\as{P}_{src} \Sigma_{\text{All \as{r}}} \as{W}(\as{r})}
\end{equation}
\noindent Which yields ${\as{mua}}_{,\as{AF},hom} \as{eta}_{\as{AF},hom}=\SI{6.81e-9}{\per\milli\meter}$, assuming the optical properties used above for the heterogeneous case.
\par

These values of ${\as{mua}}_{,\as{AF}} \as{eta}_{\as{AF}}$, whether homogeneous or heterogeneous, were used for the simulation results presented.
Specifically, by implementing their values in Equation~\ref{equ:app:R0}-\ref{equ:app:Rsim} to yield the simulated \gls{R}.
\par

\subsubsection{Fluorescence Peak Amplitude}
For the peak amplitude, we found peaks with an amplitude at least five times the noise (calculated in Appendix~\ref{sec:app:expPar:noise}) and at least \SI{1}{\second} apart. 
Considering the peak amplitude as the background-subtracted measurement, we found a mean amplitude of \SI{47.3}{\nano\ampere} or \SI{12.8}{\pico\watt\per\milli\meter\squared}.
This was for $\as{rho}=\SI{3}{\milli\meter}$ and a target depth of \SI{1.5}{\milli\meter}.
\par

Then, rewriting Equation~\ref{equ:app:Rsim} as:
\begin{equation}
    {\as{mua}}_{,target} \as{eta}_{target} \approx \frac{\as{R}_{peak}-\as{R}_{0}}
        {\as{P}_{src} \max_{\as{r}}{(\as{W}(\as{r}))}}
\end{equation}
\noindent we can find the ${\as{mua}}_{,target} \as{eta}_{target}$ by assuming that the peak maximum occurred when the target was at the location with the highest sensitivity (\gls{W}).
Using this method we found ${\as{mua}}_{,target} \as{eta}_{target}=\SI{43.3e-6}{\per\milli\meter}$.
This value, with the ones for the \gls{AF} above, was used for the simulations in this paper using Equation~\ref{equ:app:Rsim}.
\par

\subsection{Measurement of Noise}\label{sec:app:expPar:noise}
To find the noise of the \gls{I} / \gls{R}, the moving standard deviation using a \SI{1}{\second} window (with a sample rate of \SI{2}{\kilo\hertz}) was found for all experimental data, with outliers removed (defining outliers as values more than three scaled mean absolute deviations from the median), similar to the method used to find the background.
Then, the median of all moving standard deviation values was found and taken as the noise.
Using this method, we found an average measured noise at $\as{rho}=\SI{3}{\milli\meter}$, of \SI{5.29}{\nano\ampere} or \SI{1.43}{\pico\watt\per\milli\meter\squared}.
\par

Considering the peak amplitude found to be \SI{47.3}{\nano\ampere} or \SI{12.8}{\pico\watt\per\milli\meter\squared}, the \gls{SNR} at \gls{SD} $\as{rho}=\SI{3}{\milli\meter}$ was found to be \num{8.95} on average.
The \gls{sigmaRel} (Equation~\ref{equ:app:sigmaRel}) was the key parameter used to simulate noise for other distances.
Given the background value of \SI{170}{\nano\ampere} or \SI{46.1}{\pico\watt\per\milli\meter\squared}, this results in a \gls{sigmaRel} of \num{0.031}.
This value was used for the noise simulations as described in Appendix~\ref{sec:app:noise} leading to the \gls{SNR} results presented.
\par

\subsection*{Disclosures}
The authors disclose no conflicts of interest.

\subsection*{Acknowledgments}
G.B. is supported by the National Institutes of Health (NIH) award K12-GM133314, A.S. \& S.F. by award R01-EB029414, and F.I. \& M.N. by awards R21-CA246413 and R01-CA260202.
The content is solely the authors' responsibility and does not necessarily represent the official views of the awarding institutions.

\subsection*{Data, Materials, and Code Availability}
\rl{Supporting code and data can be found at:}\\
\href{https://github.com/DOIT-Lab/DOIT-Public/tree/master/DualRatio_FluorescenceFlowCytometry}{github.com/DOIT-Lab/DOIT-Public/tree/master/DualRatio\_FluorescenceFlowCytometry}
any additional supporting code and data are available from the authors upon reasonable request. 

\bibliography{refs}   

\begin{thebibliography}{10}

\bibitem{tan_vivoflowcytometry_2019}
X.~Tan, R.~Patil, P.~Bartosik, {\em et~al.}, ``In {{Vivo Flow Cytometry}} of
  {{Extremely Rare Circulating Cells}},'' {\em Scientific Reports} {\bf 9},
  3366  (2019).

\bibitem{patil_fluorescencemonitoringrare_2019}
R.~A. Patil, X.~Tan, P.~Bartosik, {\em et~al.}, ``Fluorescence monitoring of
  rare circulating tumor cell and cluster dissemination in a multiple myeloma
  xenograft model in vivo,'' {\em Journal of Biomedical Optics} {\bf 24},
  085004  (2019).

\bibitem{pace_nearinfrareddiffusevivo_2022}
J.~Pace, F.~Ivich, E.~T. Marple, {\em et~al.}, ``Near-infrared diffuse in vivo
  flow cytometry,'' {\em Journal of Biomedical Optics} {\bf 27}, 097002
  (2022).

\bibitem{zettergren_instrumentfluorescencesensing_2012}
E.~W. Zettergren, D.~Vickers, S.~K. Murthy, {\em et~al.}, ``Instrument for
  fluorescence sensing of circulating cells with diffuse light in mice in
  vivo,'' {\em Journal of Biomedical Optics} {\bf 17}, 037001  (2012).

\bibitem{sassaroli_dualslopemethodenhanced_2019}
A.~Sassaroli, G.~Blaney, and S.~Fantini, ``Dual-slope method for enhanced depth
  sensitivity in diffuse optical spectroscopy,'' {\em J. Opt. Soc. Am. A} {\bf
  36}, 1743--1761  (2019).

\bibitem{fantini_transformationalchangefield_2019}
S.~Fantini, G.~Blaney, and A.~Sassaroli, ``Transformational change in the field
  of diffuse optics: {{From}} going bananas to going nuts,'' {\em Journal of
  Innovative Optical Health Sciences} {\bf 13}, 1930013  (2019).

\bibitem{blaney_phasedualslopesfrequencydomain_2020}
G.~Blaney, A.~Sassaroli, T.~Pham, {\em et~al.}, ``Phase dual-slopes in
  frequency-domain near-infrared spectroscopy for enhanced sensitivity to brain
  tissue: {{First}} applications to human subjects,'' {\em Journal of
  Biophotonics} {\bf 13}, e201960018  (2020).

\bibitem{blaney_methodmeasuringabsolute_2022}
G.~Blaney, A.~Sassaroli, and S.~Fantini, ``Method for {{Measuring Absolute
  Optical Properties}} of {{Turbid Samples}} in a {{Standard Cuvette}},'' {\em
  Applied Sciences} {\bf 12}, 10903  (2022).

\bibitem{pang_circulatingtumourcells_2021}
S.~Pang, H.~Li, S.~Xu, {\em et~al.}, ``Circulating tumour cells at baseline and
  late phase of treatment provide prognostic value in breast cancer,'' {\em
  Scientific Reports} {\bf 11}, 13441  (2021).

\bibitem{Alix-Panabieres_ClinicalChemistry13_CirculatingTumor}
C.~{Alix-Panabi{\`e}res} and K.~Pantel, ``Circulating {{Tumor Cells}}: {{Liquid
  Biopsy}} of {{Cancer}},'' {\em Clinical Chemistry} {\bf 59}, 110--118
  (2013).
\newblock \url{https://doi.org/10.1373/clinchem.2012.194258}.

\bibitem{Mader_OncologyResearchandTreatment17_LiquidBiopsy}
S.~Mader and K.~Pantel, ``Liquid {{Biopsy}}: {{Current Status}} and {{Future
  Perspectives}},'' {\em Oncology Research and Treatment} {\bf 40}, 404--408
  (2017).
\newblock \url{https://doi.org/10.1159/000478018}.

\bibitem{Mishra_Proc.Natl.Acad.Sci.20_UltrahighthroughputMagnetic}
A.~Mishra, T.~D. Dubash, J.~F. Edd, {\em et~al.}, ``Ultrahigh-throughput
  magnetic sorting of large blood volumes for epitope-agnostic isolation of
  circulating tumor cells,'' {\em Proceedings of the National Academy of
  Sciences} {\bf 117}, 16839--16847  (2020).

\bibitem{Diamantopoulou_Nature22_MetastaticSpread}
Z.~Diamantopoulou, F.~{Castro-Giner}, F.~D. Schwab, {\em et~al.}, ``The
  metastatic spread of breast cancer accelerates during sleep,'' {\em Nature}
  {\bf 607}, 156--162  (2022).

\bibitem{williams_shorttermcirculatingtumor_2020}
A.~L. Williams, J.~E. Fitzgerald, F.~Ivich, {\em et~al.}, ``Short-{{Term
  Circulating Tumor Cell Dynamics}} in {{Mouse Xenograft Models}} and
  {{Implications}} for {{Liquid Biopsy}},'' {\em Frontiers in Oncology} {\bf
  10}  (2020).
\newblock \url{https://www.frontiersin.org/articles/10.3389/fonc.2020.601085}.

\bibitem{lee_reviewclinicaltrials_2019}
J.~Y.~K. Lee, S.~S. Cho, W.~Stummer, {\em et~al.}, ``Review of clinical trials
  in intraoperative molecular imaging during cancer surgery,'' {\em Journal of
  Biomedical Optics} {\bf 24}, 120901  (2019).

\bibitem{hernot_latestdevelopmentsmolecular_2019}
S.~Hernot, L.~van Manen, P.~Debie, {\em et~al.}, ``Latest developments in
  molecular tracers for fluorescence image-guided cancer surgery,'' {\em The
  Lancet Oncology} {\bf 20}, e354--e367  (2019).

\bibitem{niedre_prospectsfluorescencemolecular_2022}
M.~Niedre, ``Prospects for {{Fluorescence Molecular In Vivo Liquid Biopsy}} of
  {{Circulating Tumor Cells}} in {{Humans}},'' {\em Frontiers in Photonics}
  {\bf 3}  (2022).
\newblock \url{https://www.frontiersin.org/articles/10.3389/fphot.2022.910035}.

\bibitem{patil_fluorescencelabelingcirculating_2020}
R.~A. Patil, M.~Srinivasarao, M.~M. Amiji, {\em et~al.}, ``Fluorescence
  {{Labeling}} of {{Circulating Tumor Cells}} with a {{Folate Receptor-Targeted
  Molecular Probe}} for {{Diffuse In Vivo Flow Cytometry}},'' {\em Molecular
  Imaging and Biology} {\bf 22}, 1280--1289  (2020).

\bibitem{selvaraj_ultrasoundevaluationeffect_2016}
V.~Selvaraj and F.~S. Buhari, ``Ultrasound evaluation of effect of different
  degree of wrist extension on radial artery dimension at the wrist joint,''
  {\em Annals of Cardiac Anaesthesia} {\bf 19}, 63  (2016).

\bibitem{ivich_signalmeasurementconsiderations_2022}
F.~Ivich, J.~Pace, A.~L. Williams, {\em et~al.}, ``Signal and measurement
  considerations for human translation of diffuse in vivo flow cytometry,''
  {\em Journal of Biomedical Optics} {\bf 27}, 067001  (2022).

\bibitem{blaney_functionalbrainmapping_2022}
G.~Blaney, A.~Sassaroli, C.~Fernandez, {\em et~al.}, ``Functional brain mapping
  with dual-slope frequency-domain near-infrared spectroscopy,'' in {\em Neural
  {{Imaging}} and {{Sensing}}},  Q.~Luo, J.~Ding, and L.~Fu, Eds., 1194602,
  {SPIE}, ({San Francisco, CA USA})  (2022).

\bibitem{blaney_designsourcedetector_2020}
G.~Blaney, A.~Sassaroli, and S.~Fantini, ``Design of a source\textendash
  detector array for dual-slope diffuse optical imaging,'' {\em Review of
  Scientific Instruments} {\bf 91}, 093702  (2020).

\bibitem{fang_montecarlosimulation_2009}
Q.~Fang and D.~A. Boas, ``Monte {{Carlo Simulation}} of {{Photon Migration}} in
  {{3D Turbid Media Accelerated}} by {{Graphics Processing Units}},'' {\em
  Optics Express} {\bf 17}, 20178--20190  (2009).

\bibitem{Yao_Biomed.Opt.ExpressBOE18_DirectApproach}
R.~Yao, X.~Intes, and Q.~Fang, ``Direct approach to compute {{Jacobians}} for
  diffuse optical tomography using perturbation {{Monte Carlo-based}} photon
  {``replay''},'' {\em Biomedical Optics Express} {\bf 9}, 4588--4603  (2018).

\bibitem{Hartmann_Phys.Med.Biol.17_FluorescenceDetection}
C.~Hartmann, R.~Patil, C.~P. Lin, {\em et~al.}, ``Fluorescence detection,
  enumeration and characterization of single circulating cells in vivo:
  Technology, applications and future prospects,'' {\em Physics in Medicine \&
  Biology} {\bf 63}, 01TR01  (2017).

\bibitem{monici_celltissueautofluorescence_2005}
M.~Monici, ``Cell and tissue autofluorescence research and diagnostic
  applications,'' in {\em Biotechnology {{Annual Review}}},   {\bf 11},
  227--256, {Elsevier}  (2005).

\bibitem{croce_autofluorescencespectroscopyimaging_2014}
A.~C. Croce and G.~Bottiroli, ``Autofluorescence spectroscopy and imaging: A
  tool for biomedical research and diagnosis,'' {\em European Journal of
  Histochemistry} {\bf 58}  (2014).

\bibitem{hong_nearinfraredfluorophoresbiomedical_2017}
G.~Hong, A.~L. Antaris, and H.~Dai, ``Near-infrared fluorophores for biomedical
  imaging,'' {\em Nature Biomedical Engineering} {\bf 1}, 1--22  (2017).

\bibitem{gibbs_infraredfluorescenceimageguided_2012}
S.~L. Gibbs, ``Near infrared fluorescence for image-guided surgery,'' {\em
  Quantitative Imaging in Medicine and Surgery} {\bf 2}, 17787--17187  (2012).

\bibitem{thomas_identifyingnovelendogenous_2016}
G.~Thomas, M.~A. McWade, M.~E. Sanders, {\em et~al.}, ``Identifying the novel
  endogenous near-infrared fluorophore within parathyroid and other endocrine
  tissues,'' in {\em Biomedical {{Optics}} 2016 (2016), Paper {{PTu3A}}.5},
  PTu3A.5, {Optica Publishing Group}  (2016).

\bibitem{semenov_oxidationinducedautofluorescencehypothesis_2020}
A.~N. Semenov, B.~P. Yakimov, A.~A. Rubekina, {\em et~al.}, ``The
  {{Oxidation-Induced Autofluorescence Hypothesis}}: {{Red Edge Excitation}}
  and {{Implications}} for {{Metabolic Imaging}},'' {\em Molecules} {\bf 25},
  1863  (2020).

\bibitem{marmorstein_spectralprofilingautofluorescence_2002}
A.~D. Marmorstein, L.~Y. Marmorstein, H.~Sakaguchi, {\em et~al.}, ``Spectral
  {{Profiling}} of {{Autofluorescence Associated}} with {{Lipofuscin}},
  {{Bruch}}'s {{Membrane}}, and {{Sub-RPE Deposits}} in {{Normal}} and {{AMD
  Eyes}},'' {\em Investigative Ophthalmology \& Visual Science} {\bf 43},
  2435--2441  (2002).

\bibitem{huang_cutaneousmelaninexhibiting_2006}
Z.~Huang, H.~Zeng, I.~Hamzavi, {\em et~al.}, ``Cutaneous melanin exhibiting
  fluorescence emission under near-infrared light excitation,'' {\em Journal of
  Biomedical Optics} {\bf 11}, 034010  (2006).

\bibitem{wang_vivonearinfraredautofluorescence_2013}
S.~Wang, J.~Zhao, H.~Lui, {\em et~al.}, ``In vivo near-infrared
  autofluorescence imaging of pigmented skin lesions: Methods, technical
  improvements and preliminary clinical results,'' {\em Skin Research and
  Technology} {\bf 19}(1), 20--26  (2013).

\bibitem{htun_nearinfraredautofluorescenceinduced_2017}
N.~M. Htun, Y.~C. Chen, B.~Lim, {\em et~al.}, ``Near-infrared autofluorescence
  induced by intraplaque hemorrhage and heme degradation as marker for
  high-risk atherosclerotic plaques,'' {\em Nature Communications} {\bf 8}, 75
  (2017).

\bibitem{lifante_roletissuefluorescence_2020}
J.~Lifante, Y.~Shen, E.~Ximendes, {\em et~al.}, ``The role of tissue
  fluorescence in in vivo optical bioimaging,'' {\em Journal of Applied
  Physics} {\bf 128}, 171101  (2020).

\bibitem{demos_nearinfraredautofluorescenceimaging_2004}
S.~G. Demos, R.~{Gandour-Edwards}, R.~Ramsamooj, {\em et~al.}, ``Near-infrared
  autofluorescence imaging for detection of cancer,'' {\em Journal of
  Biomedical Optics} {\bf 9}, 587--592  (2004).

\bibitem{grosenick_reviewopticalbreast_2016}
D.~Grosenick, H.~Rinneberg, R.~Cubeddu, {\em et~al.}, ``Review of optical
  breast imaging and spectroscopy,'' {\em Journal of Biomedical Optics} {\bf
  21}, 091311  (2016).

\bibitem{wang_deeptissuefocalfluorescence_2012}
Y.~M. Wang, B.~Judkewitz, C.~A. DiMarzio, {\em et~al.}, ``Deep-tissue focal
  fluorescence imaging with digitally time-reversed ultrasound-encoded light,''
  {\em Nature Communications} {\bf 3}(May), 928--928  (2012).

\bibitem{klose_inversesourceproblem_2005}
A.~D. Klose, V.~Ntziachristos, and A.~H. Hielscher, ``The inverse source
  problem based on the radiative transfer equation in optical molecular
  imaging,'' {\em Journal of Computational Physics} {\bf 202}, 323--345
  (2005).

\bibitem{georgakoudi_trimodalspectroscopydetection_2002}
I.~Georgakoudi, E.~E. Sheets, M.~G. M{\"u}ller, {\em et~al.}, ``Trimodal
  spectroscopy for the detection and characterization of cervical precancers in
  vivo,'' {\em American Journal of Obstetrics and Gynecology} {\bf 186}(3),
  374--382  (2002).

\end{thebibliography}
\bibliographystyle{spiejour}   

\vspace{2ex}\noindent\textbf{Giles~Blaney} is a National Institutes of Health (NIH) Institutional Research and Academic Career Development Award (IRACDA) Postdoctoral Scholar in the Diffuse Optical Imaging of Tissue (DOIT) lab at Tufts University. He received his Ph.D. from Tufts University (Medford, MA USA) in 2022 after working in the same lab with Prof.~Sergio~Fantini as his advisor. Before that Giles received an undergraduate degree in Mechanical Engineering and Physics from Northeastern University (Boston, MA USA). His current research interests include diffuse optics and its possible applications within and outside of medical imaging.

\vspace{2ex}\noindent\textbf{Fernando~Ivich} received his B.S. and M.S. degrees in Biomedical Engineering from the University of Arizona in 2017 and 2019, respectively. He is a Ph.D. candidate in the Northeastern University Department of Bioengineering (Boston, MA USA).

\vspace{2ex}\noindent\textbf{Angelo~Sassaroli} received a Ph.D. in Physics in 2002 from the University of Electro-Communications (Tokyo, Japan). From July~2002 to August~2007, he was a Research Associate in the research group of Prof.~Sergio~Fantini at Tufts University. In September~2007 he was appointed by Tufts University as a Research Assistant Professor. His field of research is near-infrared spectroscopy and diffuse optical tomography.

\vspace{2ex}\noindent\textbf{Mark~Niedre} received his Ph.D. from the University of Toronto (Toronto, ON CA) in Medical Physics in 2004. He is a Professor of Bioengineering at Northeastern University (Boston, MA USA) and a senior member of the International Society for Optical Engineering (SPIE) and Optica.

\vspace{2ex}\noindent\textbf{Sergio~Fantini} is a Professor of Biomedical Engineering and Principal Investigator of the DOIT at Tufts University. His research activities on applying diffuse optics to biological tissues resulted in about \num{120} peer-reviewed scientific publications and \num{12} patents. He co-authored with Prof.~Irving~Bigio (Boston University, Boston, MA USA) a textbook on \say{Quantitative Biomedical Optics} published by Cambridge University Press in 2016. He is a Fellow of SPIE, Optica, and the American Institute for Medical and Biological Engineering (AIMBE).


\listoffigures
\listoftables

\end{document}